\documentclass[11pt]{article}
\usepackage{putex}
\usepackage{graphicx}
\usepackage{latexsym,amsmath,amsfonts,amssymb}
\usepackage{bbm}
\usepackage{cite}
\usepackage{mathtools}

\renewenvironment{thebibliography}[1]{%
\begin{oldthebibliography}{#1}%
\setlength\itemsep{-2mm}%
}%
{%
\end{oldthebibliography}%
}

\newcommand{\ie}{\textit{i.e.}}

\numberwithin{equation}{section}

\newcommand{\nn}{\nonumber}

\newcommand{\be}{\begin{equation}} \newcommand{\ee}{\end{equation}}
\newcommand{\bea}{\begin{equation} \begin{aligned}} \newcommand{\eea}{\end{aligned} \end{equation}}

\newcommand{\unit}{\mathbbm{1}}

\DeclareMathOperator{\Tr}{Tr}

\DeclareMathOperator{\rank}{rank}
\DeclareMathOperator*{\Res}{Res}
\DeclareMathOperator{\re}{\mathbb{R}e}
\DeclareMathOperator{\im}{\mathbb{I}m}

\begin{document}


\title{Ellipsoid partition function from Seiberg-Witten monopoles}

\authors{\textbf{Yiwen Pan and Wolfger Peelaers}\\
\vspace{10pt}
C.N. Yang Institute for Theoretical Physics, \\
Stony Brook University,\\
Stony Brook, NY, 11794}

\abstract{We study Higgs branch localization of $\mathcal N=2$ supersymmetric theories placed on compact Euclidean manifolds. We analyze the resulting localization equations in detail on the four-sphere and find that in this case the path integral is dominated by vortex-like configurations as well as singular Seiberg-Witten monopoles located at the north and south pole. The partition function is written accordingly.}

\preprint{YITP-SB-15-31}

\maketitle


{
\setcounter{tocdepth}{2}
\setlength\parskip{-0.7mm}
\tableofcontents
}


\section{Introduction}
\label{section:intro}
After it was understood in \cite{Pestun:2007rz}\footnote{See also \cite{Nekrasov:2002qd} for related earlier work.} how to apply localization techniques \cite{Witten:1988ze, Witten:1991zz} to perform exact computations of partition functions and vacuum expectation values of supersymmetric operators in supersymmetric quantum field theories defined on compact Euclidean manifolds, a wealth of exact computations in theories defined on a variety of geometries in a variety of dimensions has become available, see for example \cite{Benini:2012ui,Doroud:2012xw,Closset:2015rna,Gadde:2013dda,Benini:2013nda,Benini:2013xpa,Kapustin:2009kz,Jafferis:2010un,Hama:2010av,Hama:2011ea,Imamura:2011wg,Alday:2013lba,Alday:2012au,Assel:2014paa,Kim:2009wb,Imamura:2011su,Benini:2015noa,Hama:2012bg,Pestun:2014mja,Closset:2013sxa,Nishioka:2014zpa,Honda:2015yha,Kallen:2012cs,Hosomichi:2012ek,Kim:2012ava,Imamura:2012xg,Kim:2012gu,Terashima:2012ra,Minahan:2015jta}. Such exact, non-perturbative results can be put to excellent use in precision tests of various non-perturbative dualities, but their applications are much richer. Indeed, recently a lot of research has been conducted on interpreting and applying the wide variety of exact results available, resulting in an impressive list of both physical and mathematical developments. To name a few, the $\mathcal N = (2,2)$ $S^2$ partition function \cite{Benini:2012ui,Doroud:2012xw} computes the exact K\"ahler potential on the quantum K\"ahler moduli space of Calabi-Yau manifolds \cite{Jockers:2012dk,Gomis:2012wy,Gerchkovitz:2014gta}, the $\mathcal N = 2$ $S^3$ partition function \cite{Kapustin:2009kz,Jafferis:2010un,Hama:2010av,Hama:2011ea} is essential in the F-theorem \cite{Jafferis:2011zi}, and the partition function of four-dimensional $\mathcal N = 2$ theories placed on an ellipsoid \cite{Pestun:2007rz,Hama:2012bg} equals, for theories of class $\mathcal S$, a Liouville/Toda correlator \cite{Alday:2009aq,Wyllard:2009hg}, while for superconformal theories it also computes the K\"ahler potential on the superconformal manifold \cite{Gerchkovitz:2014gta,Gomis:2014woa}.

Localization computations are based on the observation that in the path integral of a supersymmetric theory one can add $\mathcal Q$-exact\footnote{Here $\mathcal Q$ denotes a particular supercharge of the theory, which in general is not nilpotent. Then the precise statement is that one add to the supersymmetric action $S$ a deformation term $\int \mathcal Q V,$ for some fermionic functional $V$ satisfying $\int \mathcal Q^2 V = 0.$} deformations to the action without changing the resulting partition function. For a positive semi-definite such deformation, one can then easily argue that a one-loop computation around its zeros is exact. The canonical choice of deformation term has as its zeros certain configurations involving the (bosonic) vector multiplet fields, while all matter multiplet fields are set to zero. Typically, these configurations take the form of arbitrary constant values for the vector multiplet scalars or holonomies around circles.\footnote{In the presence of homological two-cycles in the manifold, also a sum over magnetic fluxes will occur. The integration over holonomies around circles is a particular instance of the more general case which entails integration/summation over the space of flat connections. In higher-dimensional examples, also (point-like) instanton configurations will appear as zeros of the canonically chosen deformation term.} The path integral then collapses to a finite-dimensional matrix integral over this classical Coulomb branch, hence the localization based on the canonical deformation term is called Coulomb branch localization. 

Upon choosing a particular additional deformation term (or equivalently, by changing the integration contour of the auxiliary fields in complexified field space) and if certain conditions on the parameters of the theory hold, the localization locus instead consists of a finite number of discrete Higgs vacua, where matter multiplet scalars can acquire a vacuum expectation value solving the $D$-term equations, accompanied by infinite towers of non-perturbative point-like Higgs branch configurations -- \textit{e.g.}, vortices or Seiberg-Witten monopoles -- located at special points in the geometry. Such a Higgs branch localization computation was first performed in $\mathcal N = (2,2)$ theories on the two-dimensional sphere \cite{Benini:2012ui,Doroud:2012xw}, and later applied in and extended to two\cite{Closset:2015rna}, three \cite{Fujitsuka:2013fga,Benini:2013yva}, four \cite{Peelaers:2014ima} and five \cite{Pan:2014bwa} dimensional theories.\footnote{These results are closely related to the factorization results initiated in \cite{Pasquetti:2011fj}, and extended and generalized in \cite{Beem:2012mb,Hwang:2012jh,Hwang:2015wna,Taki:2013opa,Imamura:2013qxa,Yoshida:2014qwa,Nieri:2015yia}.}

In this paper, we apply the Higgs branch localization technique to $\mathcal N=2$ supersymmetric theories placed on compact Euclidean manifolds. We derive the general localization equations and subsequently study their solutions in detail on the four-sphere $S^4.$ On this geometry, the Higgs branch localization locus is given by vortex-like configurations and singular Seiberg-Witten monopoles centered at the north and south pole of $S^4$. The appearance of the latter can be understood intuitively as follows: locally around the north and south pole, the theory looks like the (anti-) topologically twisted theory with hypermultiplets. The localization locus of the latter theory is described by the Seiberg-Witten monopole equations (or their non-abelian version, \ie{}, the generalized monopole equations), which follows immediately by imposing the $D$-term equations. In the localization computation on the four-sphere, the effect of the additional deformation term is precisely to impose the $D$-term equations, while the gauge symmetry is generically broken to the maximal torus by one of the other BPS equations. Finally, we compute the resulting Higgs branch localized ellipsoid partition function. As a byproduct, we formulate a prediction for an interesting relationship between the instanton partition function and the Seiberg-Witten partition function, which capture the equivariant volume of the instanton moduli space and Seiberg-Witten moduli space respectively.

The rest of this paper is organized as follows. In section \ref{section:BPSEqns} we derive the general Higgs branch localization equations of $\mathcal N=2$ supersymmetric theories. In section \ref{section:BPSSols} we find various classes of solutions on $S^4.$ Next, in section \ref{section:computeZ}, we compute the Higgs branch localized partition function. Finally, in section \ref{section:rewrite}, we match the Coulomb branch and Higgs branch localized results in a simple example. Appendices \ref{section:sigmaandspinors} and \ref{app:N=2multiplets} summarize our conventions and recall the generalized Killing spinor equations and supersymmetry multiplets and variations. Appendix \ref{appC} studies the locally almost complex structure one can define using the Killing spinor solutions. Appendix \ref{appendix:ellipsoid} finally contains some useful specifics about the ellipsoid.

\

\emph{Note.} There has recently appeared a paper on the arXiv claiming to perform Higgs branch localization on $S^4_b$\cite{Chen:2015fta}. Our results are significantly different from theirs. Throughout the paper we will point out the major points of disagreement.


\section{BPS Equations}
\label{section:BPSEqns}
In localization computations, the path integral is localized to the zeros of a positive semi-definite deformation term $S_{\text{def.}} = \int \mathcal Q V$ satisfying $\int \mathcal Q^2 V =0.$ In this section we introduce the relevant deformation terms and derive the resulting BPS equations characterizing their zeros for $\mathcal N=2$ theories placed on manifolds admitting solutions to the generalized Killing spinor equation \eqref{Killing-spinor-equations} and auxiliary equation \eqref{auxiliary-equations}. We restrict our attention to Killing spinors satisfying the orthogonality conditions \eqref{orthocondition} guaranteeing that no scale or $U(1)_r$ transformations appear in $\mathcal Q^2.$ We also indicate how the equations simplify for the case of the ellipsoid $S^4_b.$

\subsection{Vector Multiplet}
The canonical deformation Lagrangian for the vector multiplet is given by \cite{Pestun:2007rz,Hama:2012bg}\footnote{The supersymmetry transformations are summarized in appendix \ref{app:N=2multiplets}.}
\begin{equation}
\mathcal{L}_{\text{def.}}^{\text{VM}} = \mathcal Q\Tr\left[ (\mathcal Q\lambda _{I\alpha })^\dagger \ \lambda _{I\alpha } + (\mathcal Q\tilde \lambda _I^{\dot \alpha })^\dagger \ \tilde \lambda _I^{\dot \alpha }  \right]\; \Longrightarrow\;  \mathcal{L}_{\text{def.}}^{\text{VM}}\Big|_\text{bos.} = \Tr\left[ (\mathcal Q\lambda _{I\alpha })^\dagger \ \mathcal Q\lambda _{I\alpha } + (\mathcal Q\tilde \lambda _I^{\dot \alpha })^\dagger \ \mathcal Q\tilde \lambda _I^{\dot \alpha }  \right]\;,
\end{equation}
where one considers the reality properties of various fields as in \eqref{VMreality}. We further introduce the notation ${\phi _2} \equiv \phi  - \tilde \phi  = 2\re \phi  =  - 2\re \tilde \phi $, ${\phi _1} = i\phi  + i\tilde \phi  =  - 2\im \phi  =  - 2\im \tilde \phi $. 

With the reality properties \eqref{VMreality},  $\mathcal Q\lambda_{I}$ and $\mathcal Q\tilde\lambda_I$ do not satisfy the symplectic-Majorana condition, but can be decomposed in ``real'' and ``imaginary'' pieces which do:
\begin{equation}
\mathcal Q{\lambda _I} = \re {\mathcal Q\lambda _I} + i\im {\mathcal Q\lambda _I}\;, \qquad \mathcal Q{{\tilde \lambda }_I} =\re {\mathcal Q\tilde \lambda_I} + i\im {\mathcal Q\tilde \lambda_I}\;.
\end{equation}
Explicitly, one finds
\begin{equation}
\begin{gathered}
	  \re \mathcal Q{\lambda _I} = \frac{1}{2}{\sigma ^{\mu \nu }}{\xi _I}({F_{\mu \nu }} - 4{\phi _2}({T_{\mu \nu }}+ S_{\mu\nu})) + {D_\mu }{\phi _2}{\sigma ^\mu }{{\tilde \xi }_I} \hfill \\
	  \im \mathcal Q{\lambda _I} = 2{\phi _1}\left( {{S_{\mu \nu }} - {T_{\mu \nu }}} \right){\sigma ^{\mu \nu }}{\xi _I} - \left( {{D_\mu }{\phi _1}} \right){\sigma ^\mu }{{\tilde \xi }_I} + 2{\xi _I}[\phi ,\tilde \phi ] - i{D_{IJ}}{\xi ^J} \hfill \\
	  \re \mathcal Q{{\tilde \lambda }_I} = \frac{1}{2}{{\tilde \sigma }^{\mu \nu }}{{\tilde \xi }_I}({F_{\mu \nu }} + 4{\phi _2}({{\tilde T}_{\mu \nu }}+ \tilde S_{\mu\nu})) - \left( {{D_\mu }{\phi _2}} \right){{\tilde \sigma }^\mu }{\xi _I} \hfill \\
	  \im \mathcal Q{{\tilde \lambda }_I} = 2{\phi _1}({{\tilde S}_{\mu \nu }} - {{\tilde T}_{\mu \nu }}){{\tilde \sigma }^{\mu \nu }}{{\tilde \xi }_I} - \left( {{D_\mu }{\phi _1}} \right){{\tilde \sigma }^\mu }{\xi _I} - 2{{\tilde \xi }_I}[\phi ,\tilde \phi ] - i{D_{IJ}}{{\tilde \xi }^J}\;,\hfill \\ 
	\end{gathered}  
\end{equation}
where the tensor fields $S_{\mu\nu},\tilde S_{\mu\nu}, T_{\mu\nu},$ and $\tilde T_{\mu\nu},$ are introduced in appendix \ref{app:N=2multiplets} (see \eqref{Killing-spinor-equations} and \eqref{Stensor}). The bosonic part of the deformation Lagrangian then becomes
\begin{align}
\mathcal{L}_{\text{def.}}^{\text{VM}}\Big|_\text{bos.} = \Tr\left[ (\re\mathcal Q\lambda^{I}\re\mathcal Q\lambda _{I} ) + (\im\mathcal Q\lambda^{I}\im\mathcal Q\lambda_{I} ) +  (\re\mathcal Q\tilde\lambda_{I}\re\mathcal Q\tilde\lambda^{I} ) + (\im\mathcal Q\tilde\lambda_{I}\im\mathcal Q\tilde\lambda^{I} )  \right]\;.
\end{align}
Using Fierz identities one can straightforwardly obtain (see \eqref{sum-of-squares})
\begin{align}\label{generaldeftermVM}
\mathcal{L}_{\text{def.}}^{\text{VM}}\Big|_\text{bos.} =& \Tr\left[\frac{s+\tilde s}{s\tilde s}\left((R^\mu D_\mu \phi_2)^2 + (R^\mu D_\mu \phi_1)^2 \right) - (s+\tilde s) [\phi_1,\phi_2]^2\right. \nn \\
&\qquad \left.+ \frac{1}{4s} (\re \mathcal Q\chi_{\mu\nu})^2 + \frac{1}{4s} (\im  \mathcal Q\chi_{\mu\nu})^2+\frac{1}{4\tilde s}(\re  \mathcal Q\tilde\chi_{\mu\nu})^2 + \frac{1}{4\tilde s}(\im  \mathcal Q\tilde\chi_{\mu\nu})^2\right]\;,
\end{align}
where we defined $\chi_{\mu\nu} \equiv (\xi^I \sigma_{\mu\nu}\lambda_I), \tilde\chi_{\mu\nu} \equiv (\tilde \xi_I \tilde\sigma_{\mu\nu}\tilde\lambda^I) $ and used that $\re\mathcal Q \chi_{\mu\nu} = (\xi^I \sigma_{\mu\nu} \re \lambda_I)$ and similarly for $\im\mathcal Q \chi_{\mu\nu}, \re\mathcal Q \tilde\chi_{\mu\nu}, \im\mathcal Q \tilde\chi_{\mu\nu}.$ Here we used the bilinears $s \equiv ({\xi ^I}{\xi _I}), \tilde s \equiv ({{\tilde \xi }_I}{{\tilde \xi }^I})$ and ${R^a } \equiv ({\xi ^I}{\sigma^a }{{\tilde \xi }_I}).$ One finds concretely
\begin{align}
\re\mathcal Q \chi_{\mu\nu} &=  -2 s (F_{\mu\nu}^- -4 \phi_2 (T_{\mu\nu}+S_{\mu\nu})) + 2 \left( \kappa \wedge d_A\phi_2 \right)_{\mu\nu}^- \label{ReQchimunu}\\
\im\mathcal Q \chi_{\mu\nu} &=  8s \phi_1 (T_{\mu\nu}-S_{\mu\nu}) - 2 \left(  \kappa \wedge d_A \phi_1 \right)_{\mu\nu}^-  -i \Theta_{\mu\nu}^{IJ}D_{IJ} \\
\re\mathcal Q \tilde\chi_{\mu\nu} &=   -2 \tilde s (F_{\mu\nu}^+ +4 \phi_2(\tilde T_{\mu\nu} + \tilde S_{\mu\nu})) - 2 \left( \kappa \wedge d_A\phi_2 \right)_{\mu\nu}^+  \label{ReQtildechimunu}\\
\im\mathcal Q \tilde\chi_{\mu\nu} &=  8\tilde s \phi_1(\tilde T_{\mu\nu} - \tilde S_{\mu\nu} ) - 2 \left( \kappa \wedge d_A \phi_1 \right)_{\mu\nu}^+ +i \tilde \Theta_{\mu\nu}^{IJ}D_{IJ} \;,
\end{align}
where $\Theta _{IJ}^{ab } \equiv ({\xi _I}{\sigma^{ab }}{\xi _J}), \tilde \Theta _{IJ}^{ab} \equiv ({{\tilde \xi }_I}{{\tilde \sigma }^{ab}}{{\tilde \xi }_J})$, the one-form $\kappa$ has components $\kappa_\mu = g_{\mu\nu} R^\nu,$ and $d_A$ is the gauge covariant exterior derivative. At this point, the general vector multiplet BPS equations can be read off as the arguments of the squares (with square rooted prefactors) in \eqref{generaldeftermVM}.

To perform Higgs branch localization, we add an additional deformation Lagrangian
\begin{equation}\label{defH}
\mathcal{L}_{\text{def.}}^{H_{IJ}} = \mathcal Q \Tr \left[ H^{IJ}\left( (\xi_{(I}\lambda_{J)}) - (\tilde \xi_{(I}\tilde \lambda_{J)}) \right) \right] =  \frac{1}{4}\mathcal Q \Tr \left[ H^{IJ}\left( -\frac{1}{s}\chi_{\mu\nu} \Theta^{\mu\nu}_{IJ} + \frac{1}{\tilde s} \tilde\chi_{\mu\nu} \tilde \Theta^{\mu\nu}_{IJ} \right) \right]\;,
\end{equation}
in terms of a generic adjoint valued, $SU(2)_\mathcal{R}$ triplet functional of the hypermultiplet scalars $H^{IJ},$ satisfying the reality property $(H_{IJ})^\dagger = \epsilon^{IK}\epsilon^{JL}H_{KL}.$ The second equality follows straightforwardly from a Fierz identity. The bosonic piece of the deformation Lagrangian \eqref{defH} is not positive semi-definite. However, when added to \eqref{generaldeftermVM}, the auxiliary fields $D_{IJ},$ which appear quadratically without derivatives, can be integrated out exactly by performing the Gaussian integral. Equivalently, one substitutes the $D_{IJ}$ field equation
\begin{equation}\label{EOMD}
D^{IJ} = -\frac{1}{2}H^{IJ} - \frac{4i\phi_1}{s+\tilde s}\left[\Theta^{IJ}_{\mu\nu}(T^{\mu\nu}-S^{\mu\nu})-\tilde\Theta^{IJ}_{\mu\nu}(\tilde T^{\mu\nu}-\tilde S^{\mu\nu}) \right]\;.
\end{equation}
To derive this result, we made use of the fact that $\frac{1}{s}\Theta^{\mu\nu,IJ}(\kappa\wedge \lambda)^{-}_{\mu\nu}=\frac{1}{\tilde s}\tilde\Theta^{\mu\nu,IJ}(\kappa\wedge \lambda)^{+}_{\mu\nu} $ for arbitrary one-form $\lambda$ thanks to a Fierz identity. Note also that we have effectively taken $D_{IJ}$ away from its purely imaginary integration contour. Substituting back \eqref{EOMD} in \eqref{generaldeftermVM}+\eqref{defH}, we find the following new sum of squares
\begin{align}\label{generaldeftermVM+defH}
\mathcal{L}_{\text{def.}}^{\text{VM}} + \mathcal{L}_{\text{def.}}^{H_{IJ}} \Big|_\text{bos.} =& \frac{s+\tilde s}{s\tilde s}\left((R^\mu D_\mu \phi_2)^2 + (R^\mu D_\mu \phi_1)^2 \right) - (s+\tilde s) [\phi_1,\phi_2]^2 \nn \\
&+ \frac{1}{4s} (\re \mathcal Q\chi_{\mu\nu} - \frac{1}{2}H_{IJ}\Theta^{IJ}_{\mu\nu})^2 +\frac{1}{4\tilde s}(\re  \mathcal Q\tilde\chi_{\mu\nu} + \frac{1}{2}H_{IJ}\tilde\Theta^{IJ}_{\mu\nu})^2 \nn \\
& + \frac{1}{4s} \left[ - 2 \left(  \kappa \wedge d_A \phi_1 \right)_{\mu\nu}^- + \phi_1\Theta_{\mu\nu}^{IJ}\left(w_{IJ}-\frac{1}{s+\tilde s}(s\ w_{IJ} + \tilde{s}\ \tilde w_{IJ}) \right)\right]^2\nn\\
&+ \frac{1}{4\tilde s} \left[ - 2 \left(  \kappa \wedge d_A \phi_1 \right)_{\mu\nu}^+ - \phi_1\tilde\Theta_{\mu\nu}^{IJ}\left(\tilde w_{IJ}-\frac{1}{s+\tilde s}(s\ w_{IJ} + \tilde{s}\ \tilde w_{IJ}) \right)\right]^2 \;,
\end{align}
where we used the convenient tensors
\begin{equation}
	{w_{IJ}} \equiv \frac{4}{s}\Theta _{IJ}^{\mu \nu }({T_{\mu \nu }} - {S_{\mu \nu }})\;,\qquad {\tilde w_{IJ}} \equiv  - \frac{4}{\tilde s}\tilde \Theta _{IJ}^{\mu \nu }({\tilde T_{\mu \nu }} - {\tilde S_{\mu \nu }})\;.
	\label{definition:w_IJ}
\end{equation}

\subsection{Hypermultiplet}
The canonical deformation Lagrangian for the hypermultiplet is given by
\begin{equation}
\mathcal{L}_{\text{def.}}^{\text{HM}} = \mathcal Q\left[ (\mathcal Q\psi_{\alpha A})^\dagger \psi_{\alpha A} +  (\mathcal Q\tilde\psi_{A}^{\dot\alpha })^\dagger \tilde\psi_{A}^{\dot\alpha } \right]\; \Longrightarrow\; \mathcal{L}_{\text{def.}}^{\text{VM}}\Big|_\text{bos.} =  \left[ (\mathcal Q\psi_{\alpha A})^\dagger \mathcal Q\psi_{\alpha A} +  (\mathcal Q\tilde\psi_{A}^{\dot\alpha })^\dagger \mathcal Q\tilde\psi_{A}^{\dot\alpha } \right]\;.
\end{equation}
One can split $\mathcal Q\psi_{\alpha A}$ and $\mathcal Q\tilde\psi_{A}^{\dot\alpha }$ into ``real'' and ``imaginary'' pieces, with respect to complex conjugation as in \eqref{realityHM1}, using the canonical reality properties for $q_{IA}$, but anti-canonical ones for $F_{IA}$ (see \eqref{realityHM2}, with \eqref{realityF}) \footnote{The hypermultiplet transformation rules can be found in appendix \ref{app:N=2multiplets}. The ``$\cdot$'' notation is also explained there.}
\begin{align}
\re \mathcal  Q{\psi _A} &=  - 2{\sigma ^\mu }{{\tilde \xi }^I}{D_\mu }q_{IA} - {\sigma ^\mu }{D_\mu }{{\tilde \xi }^I}q_{IA} - 2i {\xi ^I} \phi_2 \cdot  {q_{IA}}  \\
\im \mathcal  Q{\psi _A} &= - 2i\left(\xi ^I \phi_1 \cdot  q_{IA}  - {{\zeta }^{I^\prime}}F_{I^\prime A} \right) \\
\re\mathcal   Q{{\tilde \psi }_A} &=  - 2{{\tilde \sigma }^\mu }{\xi ^I}{D_\mu }q_{IA} - {{\tilde \sigma }^\mu }{D_\mu }{\xi ^I}q_{IA} + 2i\tilde \xi^I \phi_2 \cdot  q_{IA}  \\
\im \mathcal   Q{{\tilde \psi }_A} &=  -2i\left(\tilde \xi ^I\phi_1 \cdot q_{IA} - {{\tilde \zeta }^{I^\prime}}F_{I^\prime A}\right) \;,
\end{align}
which are set to zero to obtain the BPS equations. Multiplying the BPS equations following from the imaginary pieces with $\xi^K, \tilde \xi^K$ respectively, and taking their sum and difference using \eqref{zetaproperties}, one obtains
\begin{equation}
(s+\tilde s) \phi_1 \cdot q_{IA}=0\;, \qquad -\frac{1}{4}(s-\tilde s)\phi_1 \cdot q_{A}^K = \Xi^{KI^\prime}F_{I^\prime A}\;,
\end{equation}
where $\Xi^{KI^\prime} = (\xi^K\zeta^{I^\prime})=(\tilde\xi^K\tilde\zeta^{I^\prime}).$ Similarly, the real equations imply that
\begin{equation}\label{RDq-eqn}
0=2\epsilon^{JI} R^\mu D_\mu q_{IA} + 4 \left( \Theta^{IJ}_{kl}S^{kl} - \tilde\Theta^{IJ}_{kl} \tilde S^{kl} \right)q_{IA} + i \epsilon^{JI} (s-\tilde s) \phi_2 \cdot q_{IA}\;.
\end{equation}

\subsection{BPS Equations on Ellipsoid}\label{ellipsoidBPS}
For the specific case of the ellipsoid $S^4_b,$  one can use the fact that (see appendix \ref{appendix:ellipsoid})
\begin{equation}
s+\tilde s = 1\;,\qquad w_{IJ} = \tilde w_{IJ}\;,
\end{equation}
to simplify the deformation Lagrangian \eqref{generaldeftermVM+defH} to
\begin{align}\label{ellipsoidgeneraldeftermVM+defH}
\mathcal{L}_{S^4_b\text{def.}}^{\text{VM}} + \mathcal{L}_{S^4_b\text{def.}}^{H_{IJ}} \Big|_\text{bos.} =& (D_{\mu}\phi_1)^2 +  \frac{1}{s\tilde s}(R^\mu D_\mu \phi_2)^2  - [\phi_1,\phi_2]^2 \nn \\
&+ \frac{1}{4s} \left( -2 s (F_{\mu\nu}^- -4 \phi_2 (T_{\mu\nu}+S_{\mu\nu})) + 2 \left( \kappa \wedge d_A\phi_2 \right)_{\mu\nu}^- - \frac{1}{2}H_{IJ}\Theta^{IJ}_{\mu\nu}\right)^2 \nn \\
&+\frac{1}{4\tilde s}\left( -2 \tilde s (F_{\mu\nu}^+ +4 \phi_2(\tilde T_{\mu\nu} + \tilde S_{\mu\nu})) - 2 \left( \kappa \wedge d_A\phi_2 \right)_{\mu\nu}^+ + \frac{1}{2}H_{IJ}\tilde\Theta^{IJ}_{\mu\nu}\right)^2\;.
\end{align}
Here we also used that $\frac{1}{s\tilde s}(R^\mu D_\mu \phi_1)^2 +  \frac{1}{s} \left[\left(  \kappa \wedge d_A \phi_1 \right)_{\mu\nu}^- \right]^2 + \frac{1}{\tilde s} \left[ \left(  \kappa \wedge d_A \phi_1 \right)_{\mu\nu}^+\right]^2 = (D_\mu \phi_1)^2.$ The arguments of the squares in \eqref{ellipsoidgeneraldeftermVM+defH} are the ellipsoid BPS equations, which are supplemented by the $D_{IJ}$ equation of motion \eqref{EOMD}, which simplifies to
\begin{equation}\label{EOMDellipsoid}
D^{IJ} = -\frac{1}{2}H^{IJ} - i\phi_1 w^{IJ}\;.
\end{equation}
The hypermultiplet equations are given by
\begin{align}
0&= - 2{\sigma ^\mu }{{\tilde \xi }^I}{D_\mu }q_{IA} - {\sigma ^\mu }{D_\mu }{{\tilde \xi }^I}q_{IA} - 2i {\xi ^I} \phi_2 \cdot  {q_{IA}} \label{HMeqn1} \\
0&=  - 2{{\tilde \sigma }^\mu }{\xi ^I}{D_\mu }q_{IA} - {{\tilde \sigma }^\mu }{D_\mu }{\xi ^I}q_{IA} + 2i\tilde \xi^I \phi_2 \cdot  q_{IA}  \label{HMeqn2}\\
0&=\phi_1 \cdot q_{IA} \label{HMeqn3}\\
0&=F_{I^\prime A}\;.
\end{align}
Equation \eqref{RDq-eqn} also still holds.

\section{BPS Solutions}
\label{section:BPSSols}
In this section, we study the solutions to the ellipsoid BPS equations derived in subsection \ref{ellipsoidBPS}. Depending on the choice of $H_{IJ}$, we find different classes of solutions. For simplicity, we work on the round four-sphere $S^4$; the generalization to the ellipsoid is expected to be straightforward, but technically somewhat involved.

\subsection{Coulomb Branch}
Let us start by recalling the standard Coulomb branch localization locus, obtained by solving the BPS equations for $H_{IJ}=0.$ It was argued in \cite{Pestun:2007rz,Hama:2012bg} that the solution sets all hypermultiplet fields to zero, while the smooth vector multiplet solution reads 
\begin{equation}\label{CB_BPS_VM}
0=\phi_2 = A_\mu \;, \qquad \phi_1 = a\;, \qquad D_{IJ} = -i a w_{IJ}\;,
\end{equation}
for $a$ a constant, which can be chosen to lie in the Cartan subalgebra. Additionally, since $s = \sin^2 \frac{\rho}{2},\tilde s = \cos^2 \frac{\rho}{2} $ vanish at the north pole ($\rho = 0$) and the south pole ($\rho = \pi$) respectively, we see from \eqref{ellipsoidgeneraldeftermVM+defH} (with $T=\tilde T =0$ on the round four-sphere) that at the north pole the equations on the field strength relax to $F^+ = 0$ and at the south pole to $F^- = 0,$ allowing for point-like (anti-) instantons. 

Before studying the solutions that become available upon turning on $H_{IJ}$ we introduce some notation. The $A=1 \ (A=2)$ components of the hypermultiplet transform in representations $\mathfrak R\ (\bar{\mathfrak R})$ of the combined gauge and flavor group (see also appendix \ref{app:N=2multiplets}). We introduce a vector multiplet for this combined symmetry group, whose gauge group components are dynamical while its flavor group components are background, and denote its scalars as $\Phi_1, \Phi_2.$ To preserve supersymmetry, the background components need to satisfy the vector multiplet BPS equations of subsection \ref{ellipsoidBPS} (for $H_{IJ}=0$). In particular, from \eqref{CB_BPS_VM}, it is clear that one can give a vev to the background piece of $\Phi_1$ (and the background auxiliary field) which corresponds to turning on a (real) mass for the hypermultiplet.\footnote{Note that from \eqref{CB_BPS_VM} one sees that giving a vev to $\phi_2$ is not BPS on the four-sphere. Hence it is impossible to turn on the standard flat space hypermultiplet complex masses while preserving supersymmetry, contrary to what was claimed in \cite{Chen:2015fta}.} Decomposing $\mathfrak R$ into irreducible representations of the gauge group as $\mathfrak{R} = \oplus_{j} \mathcal R_j,$ we have concretely $\Phi_1|_{\mathcal R_j} = \phi_1^{(j)} + m_j, \Phi_2|_{\mathcal R_j} = \phi_2^{(j)},$ where $m_j$ is a mass for the $U(1)$ flavor symmetry carried by the hypermultiplet transforming in gauge representation $\mathcal R_j (\bar{\mathcal R}_j)$.

We choose
\begin{equation}
H_{IJ} = -\frac{\zeta}{\ell} w_{IJ} - i \sum_{j,a} T^a_{\text{adj.}}\left(q_{I1}^{(j)}\  T^a_{\mathcal R_j}\ q_{J2}^{(j)} \ +\  q_{J1}^{(j)}\  T^a_{\mathcal R_j}\ q_{I2}^{(j)} \right )\;,
\label{definition:H_IJ}
\end{equation}
where the sum runs over the irreducible gauge symmetry representations $\mathcal R_j$ and its generators $T^a_{\mathcal R_j}.$ Furthermore, $\zeta$ is a dimensionless adjoint-valued parameter defined as $\zeta  \equiv \sum_{h^a: \mathfrak u(1)} {{\zeta ^a}{h^a}},$ where the sum runs over the generators $h^a$ of $\mathfrak{u}(1)$ factors of the Lie algebra of the gauge group, and $\zeta^a$ are real parameters. It will turn out to be useful to split the $\zeta^a$ parameter in two pieces as $\zeta^a = \zeta_{\text{vac.}}^a + \zeta_{\text{SW}}^a,$ with $\zeta_{\text{vac.}}^a,\zeta_{\text{SW}}^a$ of the same sign, and define $H_{IJ}^{\text{SW}} = H_{IJ} + \frac{\zeta_{\text{vac.}}}{\ell} w_{IJ}.$

\subsection{Deformed Coulomb branch}
The deformed Coulomb branch is characterized by vanishing hypermultiplet scalars. Then, using that
\begin{equation}
d\kappa_{\mu\nu} = -8\tilde s \tilde S_{\mu\nu} - 8 s S_{\mu\nu}\;, \qquad w_{IJ} \Theta^{IJ}_{\mu\nu} = d\kappa^-_{\mu\nu}\;,\qquad w_{IJ}\tilde\Theta^{IJ}_{\mu\nu} = -d\kappa^+_{\mu\nu}\;,
\end{equation}
which are a direct consequence of the generalized Killing spinor equations on the four-sphere $S^4,$ one can write the vector multiplet equations as
\begin{equation}
\begin{gathered}
	  0 = {D_\mu }{\phi _1} = \left[ {{\phi _1},{\phi _2}} \right] = {R^\mu }{D_\mu }{\phi _2} \hfill \\
	  0 =  - 2sF_{\mu \nu }^ -  + 2\left( {\kappa  \wedge {d_A}{\phi _2}} \right)_{\mu \nu }^ -  - {\phi _2}d\kappa _{\mu \nu }^ -  + \frac{1}{2\ell}\zeta d\kappa _{\mu \nu }^ -  \hfill \\
	  0 =  - 2 \tilde sF_{\mu \nu }^ +  - 2\left( {\kappa  \wedge {d_A}{\phi _2}} \right)_{\mu \nu }^ +  + {\phi _2}d\kappa _{\mu \nu }^ +  + \frac{1}{2\ell}\zeta d\kappa _{\mu \nu }^ + \;. \hfill \\ 
	\end{gathered}
\end{equation}
Notice that on $S^4$ one can explicitly verify that
\begin{equation}
	\ell{\cos ^2}\frac{\rho }{2}d{\kappa ^ - } =  - \sin \rho {\left( {{e^4} \wedge \kappa } \right)^ - },\;\;\;\;\ell{\sin ^2}\frac{\rho }{2}d{\kappa ^ + } = \sin \rho {\left( {{e^4} \wedge \kappa } \right)^ + }\;.
\end{equation}
With this fact, it is easy to check that\footnote{Note that \cite{Chen:2015fta} set to zero all $\phi_2$ dependence. However, it is easy to verify that it is crucial to keep $\phi_2$ to have a deformed Coulomb branch configuration satisfying the Bianchi identity. As we will see, the existence of the deformed Coulomb branch solution is in turn crucial to find Higgs banch solutions.}
\begin{equation}\label{DCB_BPS_VM}
	A = \frac{1 }{3\ell}\zeta \kappa\;, \qquad {\phi _2} = \frac{1 }{6\ell}\zeta \cos \rho\;, \qquad {\phi _1} = a\;, \qquad D_{IJ} =\left( \frac{\zeta}{2\ell} - i a \right)w_{IJ}\;.
\end{equation}
for constant $a$ is a solution. Again, $a$ can be diagonalized. On top of this smooth Abelian solution, we again can have point-like (anti-) instantons located at the poles of the sphere.
        
\subsection{Higgs Branch and Seiberg-Witten Monopoles}\label{subsec:HiggsandSW}
\paragraph{Higgs-like solutions} They are characterized by the requirement that $H_{IJ}^{\text{SW}}$ vanishes. The vector multiplet equations then reduce to the deformed Coulomb branch equations of the previous subsection with deformation parameter $\zeta_{\text{vac.}},$ and have the solutions \eqref{DCB_BPS_VM}. The value of $\zeta_{\text{vac.}}$ will be fixed momentarily. In particular, the field $\phi_1$ is a diagonal, constant matrix. Its values are further constrained by the hypermultiplet equation \eqref{HMeqn3}, \ie{} $\Phi_1 \cdot q_{IA} =0.$ The combined set of equations
\begin{equation}\label{vacequations}
0=(\phi_1^{(j)} + m_j) \cdot q_{IA}^{(j)}\;, \qquad 0 = \frac{\zeta_{\text{SW}}}{\ell} w_{IJ} + i \sum_{j,a} T^a_{\text{adj.}}\left(q_{I1}^{(j)}\  T^a_{\mathcal R_j}\ q_{J2}^{(j)} \ +\  q_{J1}^{(j)}\  T^a_{\mathcal R_j}\ q_{I2}^{(j)} \right )
\end{equation}
are in fact the standard vacuum equations of an $\mathcal N=2$ supersymmetric theory in the presence of an FI-parameter. Their solutions strongly depend on the choice of gauge and matter representations of the hypermultiplets. We will be interested in cases where generic masses and generic Fayet-Iliopoulos parameters $\zeta^a$ completely break the gauge group. More precisely, we restrict ourselves to cases where the first vacuum equation in \eqref{vacequations} uniquely determines the components of $\phi_1$ in terms of the $m_j,$ and moreover where all components take distinct values, thus breaking the gauge group $G$ to its maximal torus $U(1)^{\rank G}$. The hypermultiplet scalars taking on a vacuum expectation value further break these $U(1)$s via the Higgs mechanism. One arrives at a discrete set of Higgs vacua. It is clear then that after the gauge group is broken to $U(1)^{\rank G},$ it is sufficient to analyze a $U(1)$ gauge group with a single flavor, which, up to rescaling of $U(1)$ charges, we can take to be fundamental. We will do so henceforth.

Let us consider the particular example of a $U(N_c)$ gauge group with $N_f\geq N_c$ fundamental hypermultiplets. Then it is well-known that the vacuum equations have  $\binom {N_f} {N_c}$ solutions, essentially differing by the choice of $N_c$ out of the $N_f$ hypermultiplets to acquire a vev. For positive value of $\zeta_{\text{SW}}$, and choosing the first $N_c$ hypermultiplets, one solution is given concretely as $\phi_1 = -\text{diag}(m_1,\ldots, m_{N_c})$ and $q^{ja} = \ell^{-1}\sqrt{\zeta_{\text{SW}}}\delta^{ja},$ for $j=1,\ldots, N_c,$ $q^{ja} = 0 $ for $j=N_c+1,\ldots, N_f,$ and $\tilde q_{ja} = 0,$ where $a$ denotes the gauge index and we introduced the standard notations $q\equiv q_{I=1,A=1},$ and  $\tilde q \equiv q_{I=1,A=2}$ (see also \eqref{qqtildedef}), while for negative values of $\zeta$ the $\tilde q$ get a vev. The $U(1)$ vacua can be obtained as a special case.

To complete the Higgs-like solution, we should still make sure that \eqref{HMeqn1} and \eqref{HMeqn2} are satisfied. On the round sphere, their combination \eqref{RDq-eqn} simplifies to
\begin{equation}
2\epsilon^{JI} R^\mu D_\mu q_{IA} - w^{JI}q_{IA} + i \epsilon^{JI} (s-\tilde s) \phi_2 \cdot q_{IA}=0\;,
\end{equation}
which can be decomposed in terms of the scalars $q, \tilde q^\dagger$ as
\begin{equation}\label{HMforexact}
2 R^\mu D_\mu q +\frac{i}{\ell} q + i (s-\tilde s) \ \phi_2\  q=0\;, \qquad 2 R^\mu D_\mu \tilde q^\dagger -\frac{i}{\ell} \tilde q^\dagger + i (s-\tilde s)\  \phi_2\  \tilde q^\dagger=0\;,
\end{equation}
and their complex conjugates. It is clear that in the vacuum where (only) $q$ gets a vev, these equations are solved for $\zeta_{\text{vac.}} = +6,$ while if $\tilde q$ gets a vev, one finds $\zeta_{\text{vac.}} = -6.$ It is straightforward to verify that then also \eqref{HMeqn1} and \eqref{HMeqn2} are solved.

\paragraph{Smooth ``$(m,n)$-vortex'' solutions} Let us now relax the constraint $H_{IJ}^{\text{SW}} = 0$ and study more general smooth solutions with non-zero $H_{IJ}^{\text{SW}}$\footnote{The smooth configurations of this paragraph and the singular Seiberg-Witten configurations of the next are missing in \cite{Chen:2015fta}.}. We will focus on generalizing the vacuum solutions where $\zeta$ is positive and thus $q$ acquires a vev, knowing that the case where $\zeta$ is negative and $\tilde q$ gets a vev can be treated completely similarly. Let us further denote the deformed Coulomb branch configuration for $\zeta_{\text{vac.}}$ as $A_{\text{vac.}} =  \frac{1 }{3\ell}\zeta_{\text{vac.}} \kappa,$ and $(\phi_2)_{\text{vac.}} = \frac{1}{6\ell}\zeta_{\text{vac.}} \cos \rho.$ The smooth solutions we are about to uncover carry winding around the circles parametrized by $\varphi$ and $\chi,$ and thus have their combined core at the north pole and the south pole. In fact, they resemble the standard two-dimensional vortex solutions, and as in that case, we do not have an analytic expression for the solution, but study its behavior in the far away region, and near the core. It is trivial to verify that away from the north and south pole one has the solution $\tilde q = 0$ and
\begin{equation}\label{smoothsolutionsfaraway}
A = A_{\text{vac.}} - m d\varphi - n d\chi\;, \qquad q = \frac{\sqrt{\zeta_{\text{SW}}}}{\ell} \ e^{-im\varphi - i n \chi} \;, \qquad \phi_2 = (\phi_2)_{\text{vac.}}\;,
\end{equation}
while $\phi_1$ still takes its vacuum value determined in terms of the masses. This solution is valid in the region $ \rho^2 \gg \frac{m+n}{\zeta_{\text{SW}}+2(m+n)},$ as we will derive momentarily. 

To analyze the behavior around the north pole and south pole, we study the vector and hypermultiplet equations to linear order in $\rho$ or $\pi-\rho$ respectively. The geometry is approximated by flat $\mathbb{C}^2.$ Indeed, around the north pole $\rho=0$ and with $r_1 =\ell \rho \cos\theta$ and $r_2 = \ell \rho \sin\theta$ the metric simply becomes
\begin{equation}
ds^2 = dr_1^2 + r_1^2 d\varphi^2 + dr_2^2 + r_2^2 d\chi^2\;.
\end{equation}
The hypermultiplet equations read for $\tilde q = 0$
\begin{align}
0 & = \left[D_\varphi +D_\chi + i(r_1 D_{r_1} + r_2 D_{r_2})\right] q + \mathcal O(\rho^2 q) \label{vortexlike1}\\
0 & = \left[-\frac{r_2}{r_1}D_\varphi + \frac{r_1}{r_2} D_\chi + i(-r_2 D_{r_1} + r_1 D_{r_2}) \right] q + \mathcal O(\rho^2 q) \\
0&= \left[i(-1 + \ell \phi_2) + \frac{1}{2}\left( D_\varphi +D_\chi - i(r_1 D_{r_1} + r_2 D_{r_2}) \right) \right] q  + \mathcal O(\rho^2 q)\;,
\end{align}
while the vector multiplet equations can be written as
\begingroup
\allowdisplaybreaks[1]
\begin{align}
0 & = F_{\varphi\chi} = F_{r_1r_2}\\*
{F_{\varphi {r_2}}} & = \frac{{{r_1}{r_2}}}{{2\ell }}\frac{{{r_1}}}{{r_1^2 + r_2^2}}\left[ {2{\phi _2} - \frac{\zeta }{\ell } - \frac{{4{\ell ^2}}}{{r_1^2 + r_2^2}}\left( {{r_1}{\partial _{{r_1}}} + {r_2}{\partial _{{r_2}}}} \right){\phi _2}} \right] + \mathcal O(\rho^2)\\*
{F_{\chi {r_1}}} & = \frac{{{r_1}{r_2}}}{{2\ell }}\frac{{{r_2}}}{{r_1^2 + r_2^2}}\left[ {2{\phi _2} - \frac{\zeta }{\ell } - \frac{{4{\ell ^2}}}{{r_1^2 + r_2^2}}\left( {{r_1}{\partial _{{r_1}}} + {r_2}{\partial _{{r_2}}}} \right){\phi _2}} \right] + \mathcal O(\rho^2)\\*
F_{\varphi r_1} &= r_1 \left(\frac{\phi_2}{2\ell} + \frac{\zeta}{4\ell^2}\right) - \frac{1}{2}\left(\frac{r_2}{r_1}-\frac{r_1}{r_2} \right)F_{\varphi r_2} + \mathcal O(\rho^2)\label{vortexlike2}\\*
F_{\chi r_2} &= r_2 \left(\frac{\phi_2}{2\ell} + \frac{\zeta}{4\ell^2}\right) - \frac{1}{2}\left(\frac{r_1}{r_2}-\frac{r_2}{r_1} \right)F_{\chi r_1} + \mathcal O(\rho^2)\label{vortexlike3}\;,
\end{align}
\endgroup
where we omitted terms involving  $|q|^2$ and constant times $r_i(r_1\partial_{r_1}\phi_2 +  r_2\partial_{r_2}\phi_2)$ for $i=1$ or $2,$ which assuming smoothness can only contribute at order $\rho^2$ or higher. Furthermore, we wrote the equations in such a way as to highlight the vortex-like behavior in the planes $(r_1,\varphi)$ and $(r_2,\chi)$ evident in equations \eqref{vortexlike1},\eqref{vortexlike2} and \eqref{vortexlike1},\eqref{vortexlike3} respectively if $F_{\chi r_1}=F_{\varphi r_2}=0.$ 

One finds the solution to the full set of equations \eqref{vortexlike1}-\eqref{vortexlike3} for $\rho^2 = \frac{r_1^2 + r_2^2}{\ell^2} \ll \frac{m+n}{\zeta_{\text{SW}}+2(m+n)}$ to be $\tilde q =0$ and
\begin{align}
&A - A_{\text{vac.}} = -\frac{1}{4\ell^2}\left(\frac{\zeta_{\text{SW}}}{2} + m+n \right)(r_1^2 d\varphi + r_2^2 d\chi) \;, \qquad q = B \left(r_1 e^{-i\varphi}\right)^m  \left(r_2 e^{-i\chi}\right)^n \;, \\
&\phi_2 - (\phi_2)_{\text{vac.}} = \frac{m + n}{\ell} + \frac{1}{{4{\ell^3}}}\left( -\frac{\zeta_{\text{SW}} }{{2 }} + m + n \right)(r_1^2 + r_2^2) \;,
\end{align}
for some constant $B.$ Note that $m$ and $n$ are necessarily positive and that $F_{\chi r_1}=F_{\varphi r_2}=0$ indeed. One can easily estimate the size of these smooth solutions. From \eqref{smoothsolutionsfaraway}, one can find via Stokes' theorem the vorticity of $A-A_{\text{vac.}}$ carried in the $(r_1,\varphi)$ and $(r_2,\chi)$ planes to be given by $-m$ and $-n$ respectively. Then approximating $\frac{1}{r_1}(F-F_{\text{vac.}})_{r_1\varphi}$ and $\frac{1}{r_2}(F-F_{\text{vac.}})_{r_2\chi}$ by step functions of height $-\frac{1}{2\ell^2}\left(\frac{\zeta_{\text{SW}}}{2} + m+n \right)$ on disks of radii $\epsilon_1$ and $\epsilon_2$ respectively, one easily estimates $\epsilon_1 \approx \ell \sqrt{\frac{m}{\zeta_{\text{SW}} + 2(m+n)}}$ and $\epsilon_2 \approx \ell \sqrt{\frac{n}{\zeta_{\text{SW}} + 2(m+n)}}.$ For sufficiently large values of $\zeta_{\text{SW}}$ the smooth solutions squeeze to zero size, justifying the first order approximations we made.

One can similarly analyze the asymptotic behavior near the south pole. One finds in terms of $\tilde r_1 =\ell (\pi - \rho) \cos\theta$ and $\tilde r_2 = \ell (\pi - \rho) \sin\theta$
\begin{align}
&A - A_{\text{vac.}} = -\frac{1}{4\ell^2}\left(\frac{\zeta_{\text{SW}}}{2} + m+n \right)(\tilde r_1^2 d\varphi + \tilde r_2^2 d\chi) \;, \qquad q = \tilde B \left(\tilde r_1 e^{-i\varphi}\right)^m  \left(\tilde r_2 e^{-i\chi}\right)^n \;, \\
&\phi_2 - (\phi_2)_{\text{vac.}} = -\frac{m + n}{\ell} + \frac{1}{{4{\ell^3}}}\left( \frac{\zeta_{\text{SW}} }{{2 }} - m - n \right)(\tilde r_1^2 + \tilde r_2^2) \;,
\end{align}
for $\tilde r_1^2 + \tilde r_2^2 \ll \frac{m+n}{\zeta_{\text{SW}}+2(m+n)}.$

We have constructed solutions to the BPS equations in a small neighborhood around the north pole $\rho=0$ and the south pole $\rho = \pi.$ We claim however that for $m=0$ the core of the solution (defined as the zeros of the complex scalar field $q$) in fact wraps the two-sphere $S^2_{\theta=0}$ defined by $\theta = 0$, and similarly for $n=0$ the core wraps the $\theta = \pi/2$ two-sphere $S^2_{\theta = \pi/2}$. While heuristically clear, such behavior can be argued for rigorously from the full hypermultiplet equations by starting in the core at either north or south pole and verifying that any motion along the relevant two-sphere keeps $q$ zero, but we won't do so here. In these cases we thus found a vortex-like object in $\mathbb R^2$ wrapping an $S^2.$ For generic $m,n\neq 0$ the BPS configurations are a non-trivial superposition of the $m$-vortex near $S^2_{\theta=0}$ and the $n$-vortex near $S^2_{\theta=\pi/2}$, with the core lying again on these two spheres. The cores overlap at the two intersection points of the two two-spheres, \ie{} the north pole and the south pole of $S^4.$ As we will see momentarily, there are additional point-like solutions to the BPS equations supported at these points.

The approximations made above are valid only in the limit where $\zeta_{\text{SW}}\rightarrow \infty$ and the solutions squeeze to zero size (in the planes carrying winding). For finite values of $\zeta_{\text{SW}}$ the solutions will require both finite size and curvature corrections. Nonetheless, using the BPS equations, we can deduce important properties of the solutions valid for any value of $\zeta_{\text{SW}}.$ Namely, since $q$ only vanishes in the core of the solution, we find from \eqref{HMforexact} the exact relation,
\begin{equation}
2{\iota _R}A + {\phi _2}\left( {\tilde s - s} \right)= {\ell ^{ - 1}}\left( {m + n + 1} \right)\;.\label{property:exact_relation q}
\end{equation}
Moreover, due to the compact nature of $S^4$, the winding numbers $m,n$ are not without restrictions; instead, given $\zeta_{\text{SW}}$, they are required to satisfy a certain bound, which we now derive. Consider the integral
\begin{equation}
\int {F \wedge *d\kappa }  \equiv \frac{1}{2}\int {\left( {F_{\mu \nu }^ + d\kappa _ + ^{\mu \nu } + F_{\mu \nu }^ - d\kappa _ - ^{\mu \nu }} \right)\sqrt g {d^4}x} \;.
\end{equation}
On the one hand, one can substitute the $F^{\pm}$ using the BPS equations, and obtain
\begin{equation}
	\int {F \wedge *d\kappa }  = \frac{1}{2}\int \left[ \frac{6 \phi _2}{\ell^2}\left( {s - \tilde s} \right) + \frac{1}{\ell}\left(\frac{\zeta}{\ell^2}  - \left| q \right|^2 + \left| \tilde q \right|^2\right) \right]\sqrt g d^4x \;.
	\label{eq:bound_0}
\end{equation}
On the other hand, observing that $d*d\kappa  =  - 3{\ell ^{ - 2}}\sin \rho \left( { - \sin \theta {e^1} + \cos \theta {e^2}} \right) \wedge {e^3} \wedge {e^4}$, it is easy to show that (using the fact that any complex line bundle on $S^4$ is trivial, and therefore globally $F = dA$)
\begin{equation}
	\int {F \wedge *d\kappa }  = 6{\ell ^{ - 2}}\int {\left( {{\iota _R}A} \right)\sqrt g {d^4}x} \;.
	\label{eq:bound_1}
\end{equation}
Combining the equations \eqref{eq:bound_0} and \eqref{eq:bound_1}, we have
\begin{equation}
	6{\ell ^{ - 2}}\int {\left[ {2{\iota _R}A + {\phi _2}\left( {\tilde s - s} \right)} \right]\sqrt g {d^4}x}  \leq \frac{\zeta}{\ell^3} {\Vol}\left( {{S^4}} \right)\;,
\end{equation}
where we used that on the solution $\frac{\zeta}{\ell^2}  - \left| q \right|^2 + \left| \tilde q \right|^2 \leq \frac{\zeta}{\ell^2}.$ Finally, using the exact relation (\ref{property:exact_relation q}) which implies that the integrand on the left hand side is constant on $S^4$, we obtain a bound on the winding numbers\footnote{When instead analyzing the case of negative $\zeta$ and thus non-trivial $\tilde q$ solutions, and introducing the positive winding numbers $\tilde m, \tilde n$ as in $\tilde q^\dagger \sim e^{i\tilde m \varphi +i\tilde n \chi},$ one finds the exact relation $2{\iota _R}A + {\phi _2}\left( {\tilde s - s} \right)= -{\ell ^{ - 1}}\left( {\tilde m + \tilde n + 1} \right)$ and the bound $-\tilde m - \tilde n - 1 \leqslant \frac{{\zeta }}{6}\Longrightarrow -\tilde m - \tilde n \leqslant \frac{{\zeta_{\text{SW}} }}{6}. $ }
\begin{equation}
	m + n + 1 \leqslant \frac{{\zeta }}{6} \Longrightarrow m + n  \leqslant \frac{{\zeta_{\text{SW}} }}{6}\;.
	\label{eq:bound}
\end{equation}
For finite (positive) values of $\zeta,$ only a finite number of smooth ``$(m,n)$-vortex'' solutions is supported on the four-sphere. In particular, when the bound is saturated, the scalar field $q$ (and of course $\tilde q$) vanishes and the solution is of the deformed Coulomb branch type. We thus find that upon increasing the value of $\zeta$ from zero to infinity, the Coulomb branch solutions is smoothly deformed into the deformed Coulomb branch solution, and additional smooth solutions become available each time the bound \eqref{eq:bound} is crossed. Such a picture is similar to the Higgs branch localization computations of \cite{Benini:2013yva, Peelaers:2014ima,Pan:2014bwa}.

\paragraph{Seiberg-Witten monopoles} On top of these smooth solutions we find singular solutions supported only at the north and south pole. Let us focus on the north pole first. In appendix \ref{appendix:ellipsoid} we show that there exists an integrable, self-dual complex structure $\tilde J$ which is well-defined in the region $S^4-\{{\text{south pole}}\}$. Then, taking into account that we are focusing on gauge group $G = U(1)$, we can introduce ordinary differential forms $\alpha \in \Omega^{0,0}_{\tilde J}$ and $\beta\in\Omega^{0,2}_{\tilde J}$ defined as
\begin{equation}
	\alpha  \equiv q\;, \qquad \beta  \equiv - {\tilde s^{ - 1}}{\tilde q^\dag }{\tilde \Theta _{11}}.
\end{equation}

To extract the equations describing the singular configurations, we further split off the vacuum deformed Coulomb branch solution for $\zeta = \zeta_{\text{vac.}} = 6$, \ie{} $A = A_{\text{vac.}} + a$, ${\phi _2} = (\phi _2)_{\text{vac.}} + \Delta {\phi _2}.$ Then, at the north pole ($\rho = 0$), the equations become:
\begin{align}
0 & = {\bar \partial _a}\alpha  + \bar \partial _a^*\beta  \label{eq:singular_SW_0}\\
0 & = {\Delta \phi _2}\alpha  = (\Delta{\phi _2} + 2\ell^{-1})\beta \label{eq:singular_SW_1}\\
0 & =  {F_a^{0,2}} - \frac{i}{2}\bar \alpha \beta \label{eq:singular_SW_2}\\
{F_{a}^{\tilde J}} & =- \frac{1}{4}\left[ {\frac{\zeta_\text{SW} }{{{\ell ^2}}} + \frac{{2{\Delta\phi _2}}}{\ell } - {{\left| \alpha \right|}^2} + {{\left| {\beta} \right|}^2}} \right]\tilde J\;,
\end{align}
where the superscript ${^{\tilde J}}$ denotes the component proportional to the $(1,1)_{\tilde J}$ form $\tilde J$.
By standard arguments, by combining \eqref{eq:singular_SW_0} and \eqref{eq:singular_SW_2}, it is easy to show that either $\alpha = 0$ or $\beta = 0$. As for the smooth solutions, we consider solutions with non-trivial $\alpha,$ which trivially implies that $\Delta \phi_2 = 0$ (and $\beta = 0$).\footnote{If we had split off the vacuum deformed Coulomb branch solution for $\zeta = \zeta_{\text{vac.}} = -6$, \ie{} had considered smooth solutions with non-trivial $\beta,$ then only \eqref{eq:singular_SW_1} changes. It becomes $(\Delta\phi_2 - 2\ell^{-1})\alpha = \Delta \phi_2  \beta = 0,$ thus again setting $\Delta \phi_2 = 0.$  } The equations then reduce to the standard Seiberg-Witten equations on $\mathbb{C}^2$ \cite{Witten:1994cg}, see \cite{sergeev2003vortices} for a nice introduction. Moreover, we demand that the singular solutions share the same winding numbers with the smooth solutions found above.

In the patch containing the south pole, the anti-self-dual complex structure $J$ is well-defined, see \eqref{definition:complex-structures}. It is then straightforward to derive another set of Seiberg-Witten equations at the south pole with respect to this complex structure.

Solutions $(\alpha = \alpha(z, w), \beta = 0)$ to the Seiberg-Witten equations on $\mathbb{C}^2_{z, w}$ can be constructed from complex algebraic curves, as discussed in \cite{Taubes96sw}.\footnote{We would like to thank Clifford Taubes for communication on this point.} More precisely, given a polynomial $p(z,w),$ there exists a solution to the Seiberg-Witten equations such that $\alpha(z,w)$ vanishes on the zeros of $p$ and such that near its zeros, $\alpha$ looks like the polynomial to leading order. In particular, given a polynomial $p\left( {z,w} \right) = \prod\limits_{i = 1}^D {\left( {{a}z + {b}w + {c_i}} \right)^{d_i}}, $ the preimage of zero is a collection of $D$ parallel planes intersecting the $u = az+bw$ plane at points $u=-c_i$ for $i=1,\ldots, D.$ The Seiberg-Witten solution is now in fact a multi-centered vortex solution in the $u$-plane with cores at the points $u=-c_i$ of local winding $d_i.$ Note however that the solution is not uniquely determined by the polynomial $p,$ but comes with a moduli space. 

We are looking for single-centered solutions with winding numbers matching those of the smooth solutions on top of which the singular Seiberg-Witten solution is defined. It is clear then that the relevant polynomial is given by $p(z,w)=z^m w^n.$ It will be important later to note that in particular when $m=0$ or $n=0$ we are dealing with a vortex solution in the $w$ or $z$-plane respectively.


\section{Computation of the partition function}
\label{section:computeZ}
To complete the localization computation of the partition function on $S^4,$ we need to evaluate the classical action on and the one-loop determinant of quadratic fluctuations around the BPS configurations and subsequently sum/integrate over the space of BPS solutions. Since $H_{IJ}$ is introduced through a $\mathcal Q$-exact deformation \eqref{defH}, all (appropriate) choices of $H_{IJ}$ should leave the partition function invariant. We will see in detail how this expectation works out.

Since the classical action for the hypermultiplet is $\mathcal{Q}$-exact, we only need to evaluate the Yang-Mills action \eqref{YMaction} and the Fayet-Iliopoulos action \eqref{FIaction}. Furthermore, through an index theorem, the one-loop determinants on $S^4_b$ can be computed straightforwardly \cite{Pestun:2007rz, Hama:2012bg}:
\begin{equation}\label{oneloopdets}
Z_{\text{1-loop}}^{\text{VM}}(\hat a) = \prod_{\substack{\alpha \in \mathfrak g\\\alpha\neq 0}} \Upsilon_b(i\alpha(\hat a))\;, \qquad Z_{\text{1-loop}}^{\text{HM}}(\hat a) =\prod_{w\in \mathcal R} \Upsilon_b\left(iw(\hat a)+\frac{Q}{2}\right)^{-1}\;,
\end{equation}
where $\alpha \in \mathfrak g$ are the roots of the gauge Lie algebra $\mathfrak g,$ $w\in \mathcal R$ are the weights of representation $\mathcal R$ in which the hypermultiplet transforms, and $Q = b+b^{-1},$ with $b=\sqrt{\ell/\tilde \ell}.$ Here we assumed that the (rescaled) equivariant gauge transformation parameter (see \eqref{susyalgebra}) evaluated at north and south pole are equal:
\begin{equation}\label{hata}
\hat a = \sqrt{\ell\tilde \ell}\left(2i \left( \phi \tilde s + \tilde \phi s\right) +2 i R^\mu A_\mu\right)\Big|_N=\sqrt{\ell\tilde \ell}\left(2i \left( \phi \tilde s + \tilde \phi s\right) +2 i R^\mu A_\mu\right)\Big|_S\;.
\end{equation}
Since we are working on $S^4,$ we have $\ell = \tilde \ell,$ $b=1$ and $Q=2.$ In this section we will keep using $\ell$ and $\tilde \ell$ indicating how our results are naturally generalized to the squashed sphere at the level of the partition function.

\subsection{Coulomb branch}
The classical actions evaluated on the Coulomb branch solution \eqref{CB_BPS_VM} give
\begin{equation}
S_{\text{YM}}  = \frac{8\pi^2\ell\tilde \ell}{g_{\text{YM}}^2} \Tr a^2 \;, \qquad S_{\text{FI}}  = - 16 i \pi^2 \ell \tilde \ell \Tr_{\text{FI}}a\;.
\end{equation}
The gauge equivariant parameter is easily evaluated to be $\hat a = \sqrt{\ell\tilde \ell} a,$ which can be plugged into \eqref{oneloopdets}. Taking into account the point (anti-) instantons at north and south pole, the total partition function can then be written as
\begin{equation}\label{CBpartitionfunction}
Z = \frac{1}{|\mathcal W|}\int \left(\prod_{a=1}^{\rank G} d x_a \right)\  e^{ -\frac{8\pi^2}{g_{\text{YM}}^2} \Tr x^2 + 16 i \pi^2 \sqrt{\ell\tilde \ell} \Tr_{\text{FI}}x} \ \  \frac{|Z_{\text{inst.}}(x,M,b,b^{-1},q)|^2 \ \prod_{\substack{\alpha \in \mathfrak g\\\alpha\neq 0}}\Upsilon_b(i\alpha(x))}{ \prod_j\prod_{w\in \mathcal R_j} \Upsilon_b\left(i(w(x)+M_j)+\frac{Q}{2}\right)}\;,
\end{equation}
where $Z_{\text{inst.}}(x,M,\epsilon_1,\epsilon_2,q)$ denotes the instanton partition function \cite{Nekrasov:2002qd, Nekrasov:2003rj}, we introduced $x= \sqrt{\ell\tilde \ell} a,$ and we also included the (rescaled) hypermultiplet masses $M$. The integrations are along the real line. It is furthermore relevant to mention that the poles of the instanton partition function cancel against the zeros of the vector multiplet one-loop determinant and thus the integrand only has poles originating from the hypermultiplet one-loop determinant.

\subsection{Deformed Coulomb branch}
Let us now consider the case that $\zeta\neq 0.$ On the deformed Coulomb branch configuration \eqref{DCB_BPS_VM}, the classical actions evaluate to
\begin{equation}
S_{\text{YM}}  = \frac{8\pi^2\ell\tilde \ell}{g_{\text{YM}}^2} \Tr \left(a+i\frac{\zeta Q}{12\sqrt{\ell\tilde \ell}}\right)^2 \;, \qquad S_{\text{FI}}  = - 16 i \pi^2 \ell\tilde \ell \Tr_{\text{FI}}\left(a+i\frac{\zeta Q}{12\sqrt{\ell\tilde \ell}}\right)\;.
\end{equation}
Direct evaluation of \eqref{hata} yields $\hat a = \sqrt{\ell\tilde\ell}\left(a+i\frac{\zeta Q}{12\sqrt{\ell\tilde \ell}}\right).$ Effectively, the deformed Coulomb branch thus shifts the integration contours in the matrix integral in \eqref{CBpartitionfunction} in the imaginary direction $x\rightarrow x +i\frac{\zeta Q}{12}.$ Note that to obtain these simple expression on the squashed sphere, one should apply a rescaling of $\zeta$ by a fixed function of $b$ and $b^{-1}$ (which simplifies to $1$ for $b=b^{-1}=1$).

As mentioned before, the $\zeta$-dependence is $\mathcal Q$-exact and thus should not affect the partition function. While for $\zeta=0$ one indeed recovers the standard Coulomb branch expression \eqref{CBpartitionfunction}, upon turning on $\zeta$ the integration contours are deformed and effectively shifted in the imaginary direction. The resulting integral remains equal to the original Coulomb branch integral until one crosses one of the poles of the hypermultiplet one-loop determinant. From the bound \eqref{eq:bound}, which generalizes to
\begin{equation}\label{generalbound}
mb + n b^{-1} + Q/2 \leq \frac{\zeta Q}{12}\;,
\end{equation}
one can anticipate that the positions of the poles will precisely correspond to values of $\zeta$ for which new smooth solutions become available: their contributions as well as those from the Seiberg-Witten monopoles will precisely correspond to the residues of the crossed poles.

We would like to write the partition function in terms of only the contributions of Higgs branch configurations, \ie{} we would like to find a regime of parameters for which the deformed Coulomb branch contribution vanishes in the limit $\zeta^a\rightarrow \pm \infty$. Using the asymptotic behavior of $\Upsilon_b(z)$ for large $z,$ which can be derived from the asymptotics of the double gamma function \cite{Jimbo:1996ss}, 
\begin{equation}
\log \Upsilon_b(z) \rightarrow \frac{1}{2}z(z-Q) \log\left(z(Q-z) \right) + \left(-\frac{3}{2}+\gamma \right)z(z-Q)+ \mathcal O(\log z)\;,
\end{equation}
where $\gamma$ is the Euler-Mascheroni constant, and introducing the $U(1)$ charges of the gauge representation $\mathcal R_j$ as $q^{(a)}_j \equiv w(h^a)$ for any weight $w$ of the representation $\mathcal R_j,$ we find for the leading behavior in the large $\zeta$ limit
\begin{equation}\label{asymptotics}
|\text{integrand}| \sim \exp\left[ -\frac{Q^2}{288} \sum_j \dim \mathcal R_j\  \left( \sum_a q_j^{(a)} \zeta_a \right)^2 \log \left( \sum_a q_j^{(a)} \zeta_a \right)^2  + \mathcal O(\zeta_a^2) \right]\;,
\end{equation}
which arises from the hypermultiplet one-loop determinant. The vector multiplet does not contribute to the asymptotics since it doesn't carry charge under the $U(1)$s. Similarly, only hypermultiplets transforming in representations $\mathcal R_j$ with non-zero charges $q_j^{(a)}$ contribute. Finally, note that the classical action contributes only at order $\zeta^2$. We thus conclude that if there exists a choice of $\zeta^a\rightarrow \pm \infty$ such that for a representation $\mathcal R_j$ one finds $ \sum_a q_j^{(a)} \zeta_a \rightarrow \pm \infty,$ suppression is achieved.

\subsection{Higgs branch and Seiberg-Witten Monopoles}
For finite values of $\zeta,$ one needs to take into account, apart from the contribution of the deformed Coulomb branch configuration analyzed in the previous subsection, also that of the Higgs vacua and smooth solutions, satisfying the bound \eqref{generalbound}, and the point-like Seiberg-Witten monopoles. Even though we do not possess the exact expression for the smooth solutions, it is still possible to evaluate their classical actions exactly using their behavior in the core and the BPS equations\footnote{Note that we are writing the result for the case where all $U(1)$ charges equal one. The generalization is straightforward.}:
\begin{align}\label{classicalactionsonsmooth}
& S_{\text{YM}}  = \frac{8\pi^2\ell\tilde \ell}{g_{\text{YM}}^2} \Tr \left(a_{\text{HV}}+i\frac{mb+nb^{-1}+Q/2}{\sqrt{\ell\tilde \ell}}\right)^2 \\
& S_{\text{FI}}  = - 16 i \pi^2 \ell\tilde \ell \Tr_{\text{FI}}\left(a_{\text{HV}}+i\frac{mb+nb^{-1}+Q/2}{\sqrt{\ell\tilde \ell}}\right)\;,
\end{align}
where $a_{\text{HV}}$ denotes the value of $\phi_1$ in the Higgs vacuum.

The one-loop determinants for the hypermultiplets not acquiring a vacuum expectation value and the off-diagonal vector multiplets (which in fact combine with some of the former into massive long vector multiplets through the Higgs mechanism) are straightforwardly obtained by inserting the equivariant gauge parameter \eqref{hata} evaluated on the smooth configuration,
\begin{equation}\label{hataSW}
\hat a_{\text{HV}}^{(m,n)} = \sqrt{\ell\tilde\ell} \left(a_{\text{HV}}+i\frac{mb+nb^{-1}+Q/2}{\sqrt{\ell\tilde \ell}}\right)  \;,
\end{equation}
into the one-loop determinants \eqref{oneloopdets}. The $\rank G$ hypermultiplets that do get a Higgs branch vev combine with the diagonal vector multiplets into massive long vector multiplets as well. The computation of their one-loop determinants is more subtle, as is signaled by the divergence of the hypermultiplet one-loop determinant upon inserting the equivariant gauge parameter \eqref{hataSW}. This divergence arises since we are considering a point on the Coulomb branch where these hypermultiplets are effectively massless. As explained in \cite{Doroud:2012xw}, the computation of the combined vector and hypermultiplet system is performed by removing the corresponding zero modes via a residue prescription. The total one-loop determinant can thus be written as
\begin{equation}\label{totalone-loop}
\underset{\hat a \rightarrow \hat a_{\text{HV}}^{(m,n)} }{\Res{}}\left[  Z_{\text{1-loop}}^{\text{VM}}(\hat a) Z_{\text{1-loop}}^{\text{HM}}(\hat a)\right]\;.
\end{equation}

On top of these smooth solutions, we also found point-like Seiberg-Witten monopole solutions supported at both the north pole and the south pole, see section \ref{subsec:HiggsandSW}. Their contribution, \ie{} their moduli space integral, is captured by what one may call -- in complete analogy to the (non-perturbative) $k$-instanton\footnote{If one writes the full instanton partition function as $Z_{\text{Nekr.}} = Z_{\text{pert.}} \sum_{k} q^k Z_k,$ then we mean $Z_k$ by the (non-perturbative) $k$-instanton partition function. We employ similar nomenclature for vortex partition functions and Seiberg-Witten partition functions.} and $m$-vortex partition functions -- the $p(z,w)$-Seiberg-Witten partition function $Z_{\text{SW,non-pert.}}^{\text{HV}, p(z,w)}(M,\epsilon_1,\epsilon_2,q),$ which is labeled by the particular Higgs vacuum, denoted by HV, and the complex algebraic curves $p(z,w)=z^mw^n$, and is a function of the hypermultiplet masses, the $\Omega$-deformation parameters $\epsilon_1, \epsilon_2$ and the exponentiated complexified gauge coupling $q=e^{2\pi i \tau}$. This partition function could in principle be computed by putting the $\mathcal N=2$ supersymmetric theory on $\Omega$-deformed $\mathbb R^4$ in the presence of a $\mathcal Q$-exact Fayet-Iliopoulos term such that the BPS configurations are Seiberg-Witten monopoles\footnote{This setup is quite similar to the one employed to study the two-dimensional vortex partition function, see \cite{Shadchin:2006yz}.}. The computation also requires integration over the moduli space of Seiberg-Witten solutions.

At finite values of $\zeta$, the total partition function is a sum of the contribution of the deformed Coulomb branch of the previous subsection and the Higgs vacua and smooth solutions satisfying the bound, as well as the singular Seiberg-Witten monopoles, of this subsection. The latter contribution can be written explicitly as

{\footnotesize
\begin{equation}\label{finitezetapoles}
\sum_{\substack{m,n\geq 0\\ mb + \frac{n}{b} + \frac{Q}{2} \leq \frac{\zeta Q}{12} }} e^{- \frac{8\pi^2}{g_{\text{YM}}^2} \Tr \left(\hat a_{\text{HV}}^{(m,n)} \right)^2 + 16 i \pi^2 \sqrt{\ell\tilde \ell} \Tr_{\text{FI}}\left(\hat a_{\text{HV}}^{(m,n)} \right)} \ |Z_{\text{SW,non-pert.}}^{\text{HV}, z^mw^n}(M,b,b^{-1},q)|^2 \ \underset{\hat a \rightarrow \hat a_{\text{HV}}^{(m,n)} }{\Res{}}\left[  Z_{\text{1-loop}}^{\text{VM}}(\hat a) Z_{\text{1-loop}}^{\text{HM}}(\hat a)\right]\;,
\end{equation}
\normalsize}
with $\hat a_{\text{HV}}^{(m,n)} $ defined in \eqref{hataSW} and where the modulus squared entails sending $q\rightarrow \bar q.$ It is clear that if
\begin{equation}\label{identifySW}
Z_{\text{SW,non-pert.}}^{\text{HV}, z^mw^n}(M,b,b^{-1},q) = Z_{\text{inst.}}(x_{\text{HV}} +i(mb+nb^{-1}+Q/2),M,b,b^{-1},q)\;,
\end{equation}
then \eqref{finitezetapoles} precisely contributes the residues of the crossed poles, as anticipated in the previous subsection. Unfortunately, we are not aware of an independent computation of $Z_{\text{SW,non-pert.}}^{\text{HV}, z^mw^n}(M,b,b^{-1},q),$ so instead we put forward \eqref{identifySW} as a prediction. As a particular case of \eqref{identifySW}, we find for $n=0$ that $Z_{\text{inst.}}(x_{\text{HV}} +i(mb+Q/2),M,b,b^{-1},q)$ equals the non-perturbative piece of the four-dimensional $\mathcal N=2$ supersymmetric $m$-vortex partition function in the $\Omega$-background.

\paragraph{The $\zeta \rightarrow \infty$ limit} For $\zeta \rightarrow +\infty$, we found around \eqref{asymptotics} that in favorable cases the deformed Coulomb branch contribution vanishes. Moreover, in this limit the smooth ``$(m,n)$-vortex'' solutions were found to squeeze to zero size and their winding numbers are unbounded. The path integral is thus dominated by the squeezed vortex solutions as well as the singular Seiberg-Witten solutions. We will denote their total resummed contribution in the Higgs vacuum HV rather unimaginatively as $Z_{\text{resummed}}^{(\text{HV})}.$ 

The partition function in this limit is then computed as follows:
\begin{equation}
Z = \sum_{\text{Higgs vacua HV}}\  Z_{\text{cl.}}^{(\text{HV})}\  Z_{\text{1-loop}}^{\prime\  (\text{HV})}\  Z_{\text{resummed}}^{(\text{HV})}\;.
\end{equation}
First of all, the summation runs over the finite set of all possible Higgs vacua of the theory\footnote{Recall from subsection \ref{subsec:HiggsandSW} that we restricted our attention to theories with generic masses ensuring the discrete nature of the Higgs vacua.}. The real scalar field $\phi_1$ is solved in terms of the hypermultiplet masses as $\phi_1 = a_{\text{HV}}.$ Let us also introduce $x_{\text{HV}} = \sqrt{\ell\tilde \ell} a_{\text{HV}}.$ Next, the classical actions in \eqref{classicalactionsonsmooth} provide weighting factors for $Z_{\text{resummed}}^{(\text{HV})}$ and an overall classical factor. The latter is given by
\begin{equation}
Z_{\text{cl.}}^{(\text{HV})} = \exp\left[- \frac{8\pi^2}{g_{\text{YM}}^2} \Tr \left(x_{\text{HV}}+i\frac{Q}{2}\right)^2+16 i \pi^2 \sqrt{\ell\tilde \ell} \Tr_{\text{FI}}\left(x_{\text{HV}}+i\frac{Q}{2}\right)\right]\;,
\end{equation}
while the weighting factor for the ``$(m,n)$-vortices'' and additional singular Seiberg-Witten monopoles reads
\begin{multline}\label{weighting factor}
e^{-S_{(\text{HV},m,n)}} = (q\bar q)^{\Tr\left(i b(x_{\text{HV}} + i Q/2)m - \frac{b^2}{2}m^2 \right)}\ e^{16\pi^2 i \sqrt{\ell \tilde \ell} \Tr_{\text{FI}} b m} \\ \times (q\bar q)^{\Tr\left(i  b^{-1}(x_{\text{HV}} + i Q/2)n - \frac{b^{-2}}{2}n^2  \right)}\ e^{16\pi^2 i \sqrt{\ell \tilde \ell} \Tr_{\text{FI}} b^{-1} n} \;\; (q\bar q)^{-\Tr\left(mn\right)}\;.
\end{multline}
Next, the one-loop determinant for the vectormultiplet and the chiral multiplet not acquiring a vacuum expectation value are as on the Coulomb branch, but with $x\rightarrow x_{\text{HV}}.$ The $\rank G$ chiral multiplets getting a vev produce an extra residue factor $ \left[2\pi i\  {\mathop {{\text{Res}}}\limits_{x \to 0}\Upsilon _b^{-1}\left( {ix} \right)} \right]^{\rank G}.$ Together they constitute $Z_{\text{1-loop}}^{\prime\  (\text{HV})}.$ Finally, let us give some more details on $Z_{\text{resummed}}^{(\text{HV})}.$ It can be written schematically as 
\begin{equation}
Z_{\text{resummed}}^{(\text{HV})} = \sum_{m,n\geq 0} Z_{\text{resummed}}^{(\text{HV},m,n)} = \sum_{m,n\geq 0} e^{-S_{(\text{HV},m,n)}}\  Z^{(\text{HV},m,n)}_{\text{pert.}} \ \left|Z_{\text{SW,non-pert.}}^{\text{HV}, z^m w^n}\right|^2\;,
\end{equation}
where $e^{-S_{(\text{HV},m,n)}}$ is given in \eqref{weighting factor} and $Z_{\text{SW,non-pert.}}^{\text{HV}, z^m w^n}$ was introduced below \eqref{totalone-loop}. 

Recalling the nature of the smooth solutions for $n=0,$ as a vortex of winding $m$ in the plane $\mathbb R^2$ wrapping an $S^2,$ one can identify $Z_{\text{resummed}}^{(\text{HV},m,n=0)} $ as the $S^2$-theoretic $m$-vortex partition function. Recall that we indeed already identified $Z_{\text{SW,non-pert.}}^{\text{HV}, z^m}$ as the non-perturbative piece of the $m$-vortex partition function in the $\Omega$-background. Alternatively, $Z_{\text{resummed}}^{(\text{HV},m,n=0)} $ is the $S^2$ partition function of the $m$-vortex worldvolume theory.\footnote{\label{footnote:worldvolume}Considering a four-dimensional $\mathcal N=2$ gauge theory with gauge group $U(N_c)$ and $N_f\geq N_c$ fundamental hypermultiplets, the $m$-vortex worldvolume theory is given by a two dimensional $\mathcal N=(2,2)$ gauge theory with gauge group $U(m)$ and $N_c$ fundamental chiral multiplets, $N_f-N_c$ anti-fundamental chiral multiplets and an adjoint chiral multiplet \cite{Hanany:2003hp,Hanany:2004ea}. Note that the vortex theory contains vortices itself.}  It has indeed the structure of a Higgs branch localized partition function on the two-sphere, see \cite{Benini:2012ui,Doroud:2012xw,Gomis:2014eya}. We will see an explicit example in the next section. Similar considerations are valid for $m=0.$

\section{Matching with the Coulomb branch integral}
\label{section:rewrite}
In this section we briefly show that Coulomb branch and Higgs branch localization indeed produce the same partition function for the simplest case of a $U(1)$ theory with $N_f$ fundamental hypermultiplets. The computation amounts to closing the integration contour in the Coulomb branch integral and computing the residues of the encircled poles. \footnote{The $U(N)$ generalization of this computation has been considered in \cite{Chen:2015fta}, which however performs an incorrect truncation of the sum over Young diagrams in the instanton partition function upon plugging in the position of the pole.}

We start with the Coulomb branch integral \eqref{CBpartitionfunction} specialized to the $U(1)$ case
\begin{equation}\label{U1CB}
Z^{U(1),N_f}_{S^4_b} = \int d x\  e^{ -\frac{8\pi^2}{g_{\text{YM}}^2} x^2 + 16 i \pi^2 \xi_{\text{FI}}\sqrt{\ell\tilde \ell}\  x} \ \  \frac{|Z_{\text{inst.}}(x,M,b,b^{-1},q)|^2}{ \prod_{j=1}^{N_f} \Upsilon_b\left(i(x+M_j)+\frac{Q}{2}\right)}\;.
\end{equation}
Here the masses satisfy the relation $\sum_{k=1}^{N_f} M_k = 0$ since the flavor symmetry group is $SU(N_f).$ Furthermore, the instanton partition function is given by \cite{Nekrasov:2002qd,Nekrasov:2003rj}
\begin{multline}\label{U1instanton}
Z_{\text{inst.}}(x,M,\epsilon_1,\epsilon_2,q) = \sum_{Y} q^{|Y|} \left( \prod_{j=1}^{N_f}\prod_{r=1}^{W(Y)}\prod_{s=1}^{Y_r}\left(i(x+M_j)-\frac{\epsilon_1+\epsilon_2}{2} + \epsilon_1 r + \epsilon_2 s\right)\right. \\\times\left.\frac{1}{\epsilon_2^{2|Y|}} \prod_{r,s=1}^\infty \frac{\Gamma\left(Y_r - Y_s - \frac{\epsilon_1}{\epsilon_2}(s-r+1)\right)\Gamma\left( - \frac{\epsilon_1}{\epsilon_2}(s-r) \right)}{\Gamma\left(Y_r - Y_s - \frac{\epsilon_1}{\epsilon_2}(s-r) \right)\Gamma\left( - \frac{\epsilon_1}{\epsilon_2}(s-r+1) \right)}\right)\;,
\end{multline}
as a sum over Young diagrams $Y$. Each Young diagram $Y$ encodes a non-increasing sequence of integers $(Y_1\geq Y_2 \geq \ldots \geq Y_{W(Y)}\geq Y_{W(Y)+1} = Y_{W(Y)+2}=\ldots=0),$ where $W(Y)$ is the width of the Young diagram. The total number of boxes in the diagram is denoted by $|Y|.$ The instanton counting parameter is given by $q = e^{2\pi i \tau},$ with $\tau = \frac{\theta}{2\pi} + \frac{4\pi i}{g_{\text{YM}^2}}.$ The first factor captures the contribution from the hypermultiplets, while the second factor those of the vectormultiplet. We denote them by $z_{\text{HM}}$ and $z_{\text{VM}}$ respectively. The latter can be simplified by splitting the infinite products over $r,s$ into the four regions $r,s\leq W,$ and $r\leq W, s>W,$ and $r>W,s\leq W,$ and $r>W,s> W,$ where $W\geq W(Y)$ is any integer larger than or equal to the width of the Young diagram. The last region does not contribute, while various manipulations in the other products result in
\begin{multline}\label{VMinstcontr}
z_{\text{VM}}\left(Y;x,M,b,b^{-1},q\right) \\= \left(-b^{2}\right)^{|Y|} \frac{\prod_{s=1}^{W}(1+b^2s-Y_1+Y_s)_{Y_1}}{\prod_{s=1}^W (1+b^2s)_{Y_s}\  \prod_{r,s=1}^{W} \left(1+b^2(s-r) + Y_s-Y_r\right)_{Y_r-Y_{r+1}}\  \prod_{r=1}^{W}\left(b^2(r-W-1)\right)_{Y_r}}\;,
\end{multline}
where we use the Pochhammer symbol $(y)_n = \prod_{i=0}^{n-1}(y+i)$ and already specified the $\Omega$-deformation parameters as $\epsilon_1 = b, \epsilon_2 = b^{-1}.$

As discussed around equation \eqref{asymptotics}, the contour in \eqref{U1CB} can be closed in either upper or lower-half plane. We choose to close the contour in the upper-half plane and pick up the poles at
\begin{equation}
x^{(m,n)}_j =  - {M_{j}} + imb + in{b^{ - 1}} + iQ/2\;, \qquad m,n \geq 0\;,\  j=1,\ldots, N_f\;,
\end{equation}
which are located at the zeros of the ${{\Upsilon _b}}$ functions in the denominator. Using the shift formula
\begin{align}
\Upsilon_b(x-\mu b-\nu b^{-1}) &= (-1)^{\mu \nu }\ \Upsilon_b(x)\frac{\prod_{r=0}^{\mu -1}\prod_{s=0}^{\nu -1}\left(x-(r+1)b-(s+1)b^{-1}\right)^{2}}{\prod_{r=0}^{\mu -1}\frac{\gamma(b(x-(r+1)b))}{ b^{-1+2(x-(r+1)b)b}} \ \prod_{s=0}^{\nu -1}\frac{\gamma(b^{-1}(x-(s+1)b^{-1}))}{ b^{1-2(x-(s+1)b^{-1})b^{-1}}}}\;,
\end{align}
valid for positive integers $\mu,\nu \geq 0,$ and with $\gamma(x)\equiv\frac{\Gamma(x)}{\Gamma(1-x)},$ one can straightforwardly rewrite the one-loop determinants. 

Plugging in the poles in the hypermultiplet contribution to the instanton partition function, one finds
\begin{equation}\label{HMinstcontr}
z_{\text{HM}}\left(Y;x^{(m,n)}_j,M,b,b^{-1},q\right) = \prod_{k=1}^{N_f}\prod_{r=1}^{W}\prod_{s=1}^{Y_r}\left(i M_{kj}+ b (r-(m+1)) + b^{-1} (s-(n+1))\right)\;,
\end{equation}
where we introduced $M_{kj}\equiv M_k - M_j.$ It is clear then that the contribution of diagram $Y$ vanishes if and only if it contains a box at coordinates $(\text{column},\text{row}) = (m+1,n+1).$ Note also that we trivially replaced $W(Y)$ with any integer $W \geq W(Y).$

In total we then find
\begin{equation}
Z^{U(1),N_f}_{S^4_b} = \sum_{j=1}^{N_f}\  Z_{\text{cl.}}^{(j)}\  Z_{\text{1-loop}}^{\prime(j)}\  Z_{\text{resummed}}^{(j)}\;,
\end{equation}
where 
\begin{align}
Z_{\text{cl.}}^{(j)} &= \exp\left[- \frac{8\pi^2}{g_{\text{YM}}^2} \left(-M_j+i\frac{Q}{2}\right)^2+16 i \pi^2 \sqrt{\ell\tilde \ell} \xi_{\text{FI}}\left(-M_j+i\frac{Q}{2}\right)\right]\;, \\
Z_{\text{1-loop}}^{\prime(j)}&= \frac{  2\pi i\ {\mathop {{\text{Res}}}\limits_{x \to 0} {\Upsilon _b^{-1}\left( {ix} \right)}}  }{ \prod_{\substack{k=1\\k\neq j}}^{N_f} \Upsilon_b\left(iM_{kj}+\frac{Q}{2}\right)}\;.
\end{align}
For $Z_{\text{resummed}}^{(j)}$ we find
\begin{equation}
Z_{\text{resummed}}^{(j)} = \sum_{m,n\geq 0} e^{-S_{(j,m,n)}}\  Z^{(j,m,n)}_{\text{pert.}} \ \left|Z_{\text{SW,non-pert.}}^{j, z^m w^n}\right|^2\;,
\end{equation}
where 
\begin{align}
e^{-S_{(j,m,n)}} &= (q\bar q)^{i b(-M_j + i Q/2)m - \frac{b^2}{2}m^2 }\ e^{16\pi^2 i \sqrt{\ell \tilde \ell} \xi_{\text{FI}} b m} \nn\\ &\qquad \qquad \times (q\bar q)^{i  b^{-1}(-M_j + i Q/2)n - \frac{b^{-2}}{2}n^2 }\ e^{16\pi^2 i \sqrt{\ell \tilde \ell} \xi_{\text{FI}} b^{-1} n} \;\; (q\bar q)^{-mn}\\
Z^{(j,m,n)}_{\text{pert.}}  &= \prod\limits_{k = 1}^{{N_f}} {\Bigg[ {\prod\limits_{r = 0}^{m - 1} {\frac{{\gamma \left( {b\left( {i{M_{kj}} - \left( {r + 1} \right)b} \right)} \right)}}{{{b^{ - 1 + 2\left( {i{M_{kj}} - \left( {r + 1} \right)b} \right)b}}}}} \prod\limits_{s = 0}^{n - 1} {\frac{{\gamma \left( {{b^{ - 1}}\left( {i{M_{kj}} - \left( {s + 1} \right){b^{ - 1}}} \right)} \right)}}{{{b^{1 - 2\left( {i{M_{kj}} - \left( {s + 1} \right)b^{-1}} \right){b^{ - 1}}}}}}} } } \nn \\
	& \qquad \qquad \times   { {{{\left( { - 1} \right)}^{mn}}}{ {\prod\limits_{r = 0}^{m - 1} {\prod\limits_{s = 0}^{n - 1} {{{\left( {i{M_{kj}} - \left( {r + 1} \right)b - \left( {s + 1} \right){b^{ - 1}}} \right)}^{-2}}} } } }} \Bigg]\\
Z_{\text{SW,non-pert.}}^{j, z^m w^n} & = \sum_{Y} q^{|Y|} z_{\text{VM}}\left(Y;x^{(m,n)}_j,M,b,b^{-1},q\right) z_{\text{HM}}\left(Y;x^{(m,n)}_j,M,b,b^{-1},q\right)\;,
\end{align}
where we should insert \eqref{VMinstcontr} and \eqref{HMinstcontr} in the last line. As discussed below \eqref{HMinstcontr}, the sum over Young diagrams $Y$ is effectively truncated to those diagrams whose shape is such that they do not contain a box with coordinates $(\text{column},\text{row}) = (m+1,n+1).$

\paragraph{The Special case of $n=0.$} In the previous section we have argued that for $n=0,$ $Z_{\text{resummed}}^{(j,m,n=0)} $ should equal the $S^2$-theoretic $m$-vortex partition function, or equivalently, the $S^2$ partition function of the $m$-vortex worldvolume theory. Let us see how these expectations are realized in the concrete example at hand. We find
\begin{equation}\label{tocompare}
Z_{\text{resummed}}^{(j,m,n=0)}  =  e^{-S_{(j,m,n=0)}}\  Z^{(j,m,n=0)}_{\text{pert.}} \ \left|Z_{\text{SW,non-pert.}}^{j, z^m}\right|^2\;,
\end{equation}
with
\begin{align}
e^{-S_{(j,m,n=0)}} &= (q\bar q)^{i b(-M_j + i Q/2)m - \frac{b^2}{2}m^2 }\ e^{16\pi^2 i \sqrt{\ell \tilde \ell} \xi_{\text{FI}} b m}\\
Z^{(j,m,n=0)}_{\text{pert.}}  &= \left(b^{-2N_f}\right)^{ib(-M_j+iQ/2)m - \frac{b^2}{2}m^2 } \ \prod\limits_{r = 0}^{m - 1} \frac{\gamma \left(  - \left( {r + 1} \right)b^2  \right)}{\prod_{\substack{k = 1\\k\neq j}}^{N_f} {\gamma \left( 1-{b\left( {i{M_{kj}} - \left( {r + 1} \right)b} \right)} \right)}   }\;,
\end{align}
after some straightforward manipulations and using in the second line that the masses sum to zero. Before writing down $Z_{\text{SW,non-pert.}}^{j, z^m}$ we should first remark that for $n=0$ only Young diagrams $Y$ not containing a box with coordinates $(\text{column},\text{row}) = (m+1,1),$ \ie{}, satisfying $W(Y) \leq m,$ have non-vanishing contributions. We can thus use $m$ as $W$ in \eqref{VMinstcontr} and \eqref{HMinstcontr} and obtain
\begin{multline}
Z_{\text{SW,non-pert.}}^{j, z^m} = \\ \sum_{Y}^\prime \left(-b^{2-N_f} q\right)^{|Y|} \frac{\prod_{s=1}^{m}(1+b^2s-Y_1+Y_s)_{Y_1}\ \prod_{\substack{k=1\\k\neq j}}^{N_f}\prod_{r=1}^{m}\prod_{s=1}^{Y_r}\left(i b M_{kj}+ b^2 (r-m-1) +  (s-1)\right)}{\prod_{s=1}^m (1+b^2s)_{Y_s}\  \prod_{r,s=1}^{m} \left(1+b^2(s-r) + Y_s-Y_r\right)_{Y_r-Y_{r+1}}},
\end{multline}
where the prime indicates the restriction $W(Y)\leq m.$

Expression \eqref{tocompare} should be compared to the $S^2$ partition function of an $\mathcal N=(2,2)$ supersymmetric $U(m)$ gauge theory with $N_c=1$ fundamental chiral multiplets, $N_f-N_c=N_f-1$ anti-fundamental chiral multiplets and an adjoint chiral multiplet (see footnote \ref{footnote:worldvolume}). In the conformal case, $N_f = 2N_c = 2,$ this partition function was computed in \cite{Gomis:2014eya} (see their section 2.3.1) and we find almost\footnote{A discrepancy arises from the factor $\left(b^{-4}\right)^{ib(-M_j+iQ/2)m - \frac{b^2}{2}m^2 }$ in the perturbative part.} perfect agreement upon identifying
\begin{equation}
\frac{4\pi}{g_{YM}^2} = \xi_{\text{FI}}^{(2d)}\;,\qquad \theta_{4d} = \theta_{2d} + (m-1)\pi\;, \qquad  \xi_{\text{FI}}^{(4d)}=0\;,
\end{equation}
and 
\begin{equation}
-ib^2 = m_X\;, \qquad b M_j - ib^2 -i/2 = m\;, \qquad -b M_{k(\neq j)} + \frac{i}{2} = \tilde m \;.
\end{equation}
where we denoted the masses in the two-dimensional theory as $m$ for the fundamental chiral multiplet, $\tilde m$ for the anti-fundamental chiral multiplet, and $m_X$ for the adjoint chiral multiplet. In particular we find that $Z_{\text{SW,non-pert.}}^{j, z^m}$ precisely equals the vortex partition function of the two-dimensional theory.


\section*{Acknowledgements}
We would like to thank Armen Sergeev and Clifford Taubes for communication on Seiberg-Witten solutions. We are grateful to Jaume Gomis, Leonardo Rastelli and Martin Rocek for comments on the manuscript. The work of Y.P. and W.P. is supported in part by NSF Grant PHY-1316617.

\appendix

\section{Sigma matrices and Spinors}
\label{section:sigmaandspinors}

In this appendix we review our conventions, following \cite{Hama:2012bg}, for spinors and $\sigma$-matrices.

\subsection{Spinors}
The spinor indices $\alpha, \beta,\ldots = 1,2$ and $\dot \alpha, \dot \beta, \ldots = 1, 2$ of (anti-)chiral spinors $\psi_\alpha$, ${\tilde \psi ^{\dot \alpha }}$, are raised and lowered by antisymmetric $\epsilon$-tensors, which take values $\epsilon^{12} = - \epsilon_{12} = \epsilon^{\dot{1}\dot{2}} = - \epsilon_{\dot 1 \dot 2} = 1,$ as follows:
\begin{equation}
	{\psi ^\alpha } \equiv {\epsilon ^{\alpha \beta }}{\psi _\beta }\;,\qquad {{\tilde \psi }_{\dot \alpha }} = {\epsilon _{\dot \alpha \dot \beta }}{{\tilde \psi }^{\dot \beta }}\;.
\end{equation}
The spinor product is denoted by parenthesis $(\ )$ and defined as
\begin{equation}
	\left( {\psi \chi } \right) \equiv {\psi ^\alpha }{\chi _\alpha }\;,\qquad(\tilde \psi \tilde \chi ) \equiv {{\tilde \psi }_{\dot \alpha }}{{\tilde \chi }^{\dot \alpha }}\;.
\end{equation}
A symplectic-Majorana spinor $\psi_I$ or $\tilde \psi_I$ is a doublet of chiral or anti-chiral spinors satisfying
\begin{equation}
	{({\psi _{I\alpha }})^\dag } = {\epsilon ^{IJ}}{\epsilon ^{\alpha \beta }}{\psi _{J\beta }}\;,\qquad{({\tilde \psi _{I\dot \alpha }})^\dag } = {\epsilon ^{IJ}}{\epsilon ^{\dot \alpha \dot \beta }}{\psi _{J\dot \beta }}\;.
\end{equation}
The doublet indices $I, J, K, \ldots = 1,2$ are raised and lowered by $\varepsilon^{IJ}, \varepsilon_{IJ}$, with ${\varepsilon^{12}} =  - {\varepsilon _{12}} = 1$, as ${X^I} = {\varepsilon ^{IJ}}{X_J}$, ${X_I} = {\varepsilon _{IJ}}{X^J}$. Note that $({\psi ^I}{\psi _I}) = {\left( {{\psi _{I\alpha }}} \right)^\dag }{\psi _{I\alpha }} $, and $({{\tilde \psi }_I}{{\tilde \psi }^I}) = {(\tilde \psi _I^{\dot \alpha })^\dag }\tilde \psi _I^{\dot \alpha }$ are semi-positive products.

Finally, given symplectic-Majorana spinors $\xi_I$ or $\tilde \xi_I$, one can define several useful bilinears, including
\begin{equation}
s \equiv ({\xi ^I}{\xi _I})\;,\qquad \tilde s \equiv ({{\tilde \xi }_I}{{\tilde \xi }^I})\;,\qquad {R^a } \equiv ({\xi ^I}{\sigma^a }{{\tilde \xi }_I})\;,\qquad  \Theta _{IJ}^{ab } \equiv ({\xi _I}{\sigma^{ab }}{\xi _J})\;, \qquad \tilde \Theta _{IJ}^{ab} \equiv ({{\tilde \xi }_I}{{\tilde \sigma }^{ab}}{{\tilde \xi }_J})\;, \label{bilinearsofKS}
\end{equation}
where we used the $\sigma$-matrices introduced in the next subsection.

\subsection{$\sigma$-matrices and Fierz identities}
We introduce the $\sigma$-matrices $(\sigma^a)_{\alpha\dot\alpha}, (\tilde\sigma^a)^{\dot\alpha\alpha}$, for $a=1,\ldots,4$, as
\begin{equation}
	{\sigma^a } = \left( { - i{\tau _1}, - i{\tau _2}, - i{\tau _3},{\unit_{2 \times 2}}} \right),\;\;\;\;{{\tilde \sigma }^a } = \left( {i{\tau _1},i{\tau _2},i{\tau _3},{\unit_{2 \times 2}}} \right).
\end{equation}
They satisfy the defining anti-commutation relations ${\sigma^a }{{\tilde \sigma }^b } + {\sigma^b }{{\tilde \sigma }^a } = 2{\delta^{ab }}\unit_{2 \times 2}$, and ${{\tilde \sigma }^a }{\sigma^b } + {{\tilde \sigma }^b }{\sigma^a } = 2{\delta^{ab}}\unit_{2 \times 2}$. We further introduce the anti self-dual $\sigma^{ab} = \frac{1}{2}(\sigma^a\tilde\sigma^b - \sigma^b\tilde\sigma^a),$ and self-dual $\tilde\sigma^{ab} = \frac{1}{2}(\tilde\sigma^a\sigma^b - \tilde\sigma^b\sigma^a).$

The basic Fierz identities are (for commuting spinors $\psi_i, \tilde\psi_j$)
\begin{align}
	{\sigma _\mu }{{\tilde \psi }_1}({{\tilde \psi }_2}{{\tilde \sigma }^\mu }{\psi _3}) = 2{\psi _3}({{\tilde \psi }_2}{{\tilde \psi }_1})\;,\qquad &{{\tilde \sigma }^\mu }{\psi _1}({\psi _2}{\sigma _\mu }{{\tilde \psi }_3}) = 2{{\tilde \psi }_3}\left( {{\psi _2}{\psi _1}} \right)\;,\\
{\psi _1}\left( {{\psi _2}{\psi _3}} \right) - \frac{1}{4}{\sigma _{ab }}{\psi _1}\left( {{\psi _2}{\sigma ^{ab }}{\psi _3}} \right) = 2{\psi _3}\left( {{\psi _2}{\psi _1}} \right)\;,\qquad &{{\tilde \psi }_1}({{\tilde \psi }_2}{{\tilde \psi }_3}) - \frac{1}{4}{\sigma _{ab }}{{\tilde \psi }_1}({{\tilde \psi }_2}{{\tilde \sigma }^{ab}}{{\tilde \psi }_3}) = 2{{\tilde \psi }_3}({{\tilde \psi }_2}{{\tilde \psi }_1})\;.
\end{align}
Combining the Fierz identities in the second line, one can replace spinor products with tensor products of bilinears,
\begin{equation}
	\left\{ \begin{gathered}
	  \left( {{\psi ^I}{\psi _I}} \right) = \frac{1}{s}({\xi ^I}{\psi _I})({\xi ^J}{\psi _J}) + \frac{1}{{4s}}({\xi ^I}{\sigma ^{ab }}{\psi _I})({\xi ^J}{\sigma _{ab }}{\psi _J}) \hfill \\
	  ( {{{\tilde \psi }_I}{{\tilde \psi }^I}} ) = \frac{1}{{\tilde s}}({{\tilde \xi }_I}{{\tilde \psi }^I})({{\tilde \xi }_J}{{\tilde \psi }^J}) + \frac{1}{{4\tilde s}}({{\tilde \xi }_I}{{\tilde \sigma }^{ab }}{{\tilde \psi }^I})({{\tilde \xi }_J}{{\tilde \sigma }_{ab }}{{\tilde \psi }^J})\;. \hfill \\ 
	\end{gathered}  \right.
	\label{sum-of-squares}
\end{equation}
These identities will be useful in section \ref{section:BPSEqns} for rewriting the $\mathcal Q$-exact deformation terms. 


\section{$\mathcal{N} = 2$ Vector Multiplet and Hypermultiplet}
\label{app:N=2multiplets}
\paragraph{$\mathcal N=2$ Killing spinors on Eculidean four-manifolds.}
As discussed in \cite{Hama:2012bg} (see also \cite{Bawane:2014uka, Pestun:2014mja}), four-dimensional $\mathcal{N} = 2$ supersymmetric theories can be placed on compact Euclidean manifolds with metric $g_{\mu\nu}$ if one can find symplectic-Majorana spinors $\xi_I$ and $\tilde \xi_I$ solving the generalized Killing spinor equations
\begin{equation}
	\begin{gathered}
	  {D_\mu }{\xi _I} =  - {T^{\lambda \rho }}{\sigma _{\lambda \rho }}{\sigma _\mu }{{\tilde \xi }_I} - i{\sigma _\mu }{{\tilde \xi '}_I} \hfill \\
	  {D_\mu }{{\tilde \xi }_I} =  - {{\tilde T}^{\lambda \rho }}{{\tilde \sigma }_{\lambda \rho }}{{\tilde \sigma }_\mu }{\xi _I} - i{{\tilde \sigma }_\mu }{{ \xi }_I^\prime} \hfill \\ 
	\end{gathered} 
	\label{Killing-spinor-equations}
\end{equation}
and the auxiliary equations
\begin{equation}
	\begin{gathered}
	  {\sigma ^\mu }{{\tilde \sigma }^\nu }{D_\mu }{D_\nu }{\xi _I} + 4{D_\lambda }{T_{\mu \nu }}{\sigma ^{\mu \nu }}{\sigma ^\lambda }{{\tilde \xi }_I} = M{\xi _I} \hfill \\
	  {{\tilde \sigma }^\mu }{\sigma ^\nu }{D_\mu }{D_\nu }{{\tilde \xi }_I} + 4{D_\lambda }{{\tilde T}_{\mu \nu }}{{\tilde \sigma }^{\mu \nu }}{{\tilde \sigma }^\lambda }{\xi _I} = M{{\tilde \xi }_I} \;.\hfill \\ 
	\end{gathered} 
	\label{auxiliary-equations}
\end{equation}
Here $I,J,\ldots=1,2$ denote $SU(2)_\mathcal{R}$ indices. The generalized Killing spinor equations and auxiliary equations contain the real (anti-) self-dual tensor background fields $T_{\mu\nu},\tilde T_{\mu\nu},$ and the scalar background field $M.$ Moreover, the covariant derivatives contain an $SU(2)_\mathcal{R}$ background gauge field $(V_{\mu})^I_{\phantom{I}J}$. The spinors $\xi_I^\prime, \tilde \xi_I^\prime $ are arbitrary anti-symplectic-Majorana spinors. These primed spinors and the background fields $T,\tilde T, M,$ and $V$ are part of the freedom in solving the equations. We take $\xi_I$ and $\tilde \xi_I$ to be bosonic spinors satisfying the above equations and denote the corresponding supercharge as $\mathcal Q.$ 

\paragraph{Vector Multiplet.}
An off-shell $\mathcal{N} = 2$ vector multiplet contains the gauge field $A_\mu,$ complex scalars  $\phi, \tilde \phi,$  (anti-) chiral symplectic-Majorana spinors $\lambda_I$ and $\tilde \lambda_I,$ and an $SU(2)_\mathcal{R}$ triplet of auxiliary fields $D_{(IJ)}.$ Their transformation rules are given by \cite{Hama:2012bg}
\begin{equation}
	\left\{ \begin{gathered}
	  {\mathcal Q}{A_\mu } = i({\xi ^I}{\sigma _\mu }{{\tilde \lambda }_I}) - i({{\tilde \xi }^I}{{\tilde \sigma }_\mu }{\lambda _I}) \hfill \\
	  {\mathcal Q}\phi  =  - i\left( {{\xi ^I}{\lambda _I}} \right) \hfill \\
	  {\mathcal Q}\tilde \phi  =  + i({{\tilde \xi }^I}{{\tilde \lambda }_I}) \hfill \\
	  {\mathcal Q}{\lambda _I} = \frac{1}{2}{\sigma ^{\mu \nu }}{\xi _I}( {{F_{\mu \nu }} + 8\tilde \phi {T_{\mu \nu }}} ) + 2\left( {{D_\mu }\phi } \right){\sigma ^\mu }{{\tilde \xi }_I} + \phi {\sigma ^\mu }{D_\mu }{{\tilde \xi }_I} + 2i{\xi _I}[\phi ,\tilde \phi ] + {D_{IJ}}{\xi ^J} \hfill \\
	  {\mathcal Q}{{\tilde \lambda }_I} = \frac{1}{2}{{\tilde \sigma }^{\mu \nu }}{{\tilde \xi }_I}( {{F_{\mu \nu }} + 8\phi {{\tilde T}_{\mu \nu }}}) + 2({D_\mu }\tilde \phi ){{\tilde \sigma }^\mu }{\xi _I} + \tilde \phi {{\tilde \sigma }^\mu }{D_\mu }{\xi _I} - 2i{{\tilde \xi }_I}[\phi ,\tilde \phi ] + {D_{IJ}}{{\tilde \xi }^J} \hfill \\
	  {\mathcal Q}{D_{IJ}} =  - ({{\tilde \xi }_I}{{\tilde \sigma }^\mu }{D_\mu }{\lambda _J}) + i({\xi _I}{\sigma ^\mu }{D_\mu }{{\tilde \lambda }_J}) - 2[\phi ,({{\tilde \xi }_I}{{\tilde \lambda }_J})] + 2[\tilde \phi ,\left( {{\xi _I}{\lambda _J}} \right)] + \left( {I \leftrightarrow J} \right) \;. \hfill \\ 
	\end{gathered}  \right.
	\label{vector-multiplet-transformation}
\end{equation}

The supersymmetric Yang-Mills action is given by \cite{Hama:2012bg}
\begin{align}
&S_{\text{YM}} =\frac{1}{g_{\text{YM}}^2} \int d^4 x \sqrt{g} \Tr  \left[ \frac{1}{2}F_{\mu\nu} F^{\mu\nu} + 16 F_{\mu\nu}(\tilde \phi T^{\mu\nu} + \phi \tilde T^{\mu\nu}) + 64 \tilde \phi^2 T_{\mu\nu} T^{\mu\nu} + 64 \phi^2 \tilde T_{\mu\nu} \tilde T^{\mu\nu}  \right. \nn \\
&\left.  - \frac{1}{2}D^{IJ}D_{IJ} -4D_\mu\tilde\phi D^\mu \phi + 2M\tilde \phi \phi + 4[\phi,\tilde\phi]^2 - 2i (\lambda^I\sigma^\mu D_\mu\tilde\lambda_I)-2(\lambda^I[\tilde\phi,\lambda_I]) + 2(\tilde\lambda^I[\phi,\tilde\lambda_I]) \right]\;,\label{YMaction}
\end{align}
which is positive definite upon imposing the reality properties 
\begin{equation}\label{VMreality}
A_\mu^\dagger = A_\mu\;, \qquad \phi ^\dag  =  - \tilde \phi\;, \qquad {\left( {{D_{IJ}}} \right)^\dag } =  - {D^{IJ}}\;.
\end{equation}
on the bosonic fields, while one maintains the symplectic-Majorana nature of $\lambda^I, \tilde \lambda^I.$ If the gauge group contains a $U(1)$ factor, one can also introduce a Fayet-Iliopoulos term. Introducing an $SU(2)_{\mathcal R}$ triplet background field $w^{IJ}$ satisfying
\begin{equation}
w^{IJ}\xi_J = -2i \xi^{\prime I} + 2 T^{\mu\nu}\sigma_{\mu\nu} \xi^I\;, \qquad w^{IJ}\tilde\xi_J = -2i \tilde\xi^{\prime I} + 2  \tilde T^{\mu\nu}\tilde\sigma_{\mu\nu} \xi^I\;,
\end{equation}
one can write the invariant action \cite{Hama:2012bg}
\begin{equation}\label{FIaction}
S_{\text{FI}} = \int d^4 x \sqrt{g} \Tr_{\text{FI}}\left[w^{IJ}D_{IJ} - M(\phi + \tilde \phi) - 64 \phi T^{\mu\nu}T_{\mu\nu} - 64 \tilde\phi \tilde T^{\mu\nu}\tilde T_{\mu\nu} - 8 F^{\mu\nu}(T_{\mu\nu}+\tilde T_{\mu\nu}) \right]\;,
\end{equation}
where $\Tr_{\text{FI}}$ denotes a trace that weighs each $U(1)$ factor in the gauge group with its own Fayet-Iliopoulos parameter $\xi_{\text{FI}}.$

\paragraph{Hypermultiplet.}
An off-shell $\mathcal N=2$ hypermultiplet\footnote{The multiplet is off-shell with respect to the particularly chosen supercharge $\mathcal Q$ corresponding to the Killing spinors $\xi_I, \tilde \xi_I.$} consists of scalar fields $q_{IA}$, the fermions ${\psi_A}, {\tilde \psi_A},$ and the auxiliary fields $F_{I^\prime A}$. Here $A, B, C, \ldots = 1, 2$ denote $USp(2)$ indices (which is broken to the Cartan upon gauging), and $I', J', ... = 1, 2$ are $SU(2)_{\mathcal{R}'}$ indices. Furthermore $\psi_A$ and $\tilde \psi_A$ are $\Omega$-symplectic-Majorana spinors,
\begin{equation}\label{realityHM1}
(\psi_{\alpha A})^\dagger = \epsilon^{\alpha\beta}\Omega^{AB}\psi_{\beta B}\;, \qquad (\tilde \psi_{\dot \alpha A})^\dagger = \epsilon^{\dot\alpha\dot\beta}\Omega^{AB}\tilde\psi_{\dot\beta B}\;,
\end{equation}
while $q$ and $F$ (canonically\footnote{The reality property of $F_{IA}$ will be changed in \eqref{realityF} to $\overline{\left( {F_{I^\prime A}} \right) } = -{\Omega^{AB}}{\epsilon^{I^\prime J^\prime}}F_{J^\prime B}$ to ensure a positive definite action. }) have reality properties
\begin{equation}\label{realityHM2}
	\overline{\left( {q_{IA}} \right)} = {\Omega^{AB}}{\epsilon^{IJ}}q_{JB}\;,\qquad \overline{\left( {F_{I^\prime A}} \right) } = {\Omega^{AB}}{\epsilon^{I^\prime J^\prime}}F_{J^\prime B}\;,
\end{equation}
where ${\Omega _{12}} =  - {\Omega ^{12}} = -1$ is the symplectic form of $USp(2).$ Note that the reality property of $q$ implies it can be written as
\begin{equation}\label{qqtildedef}
	{q_{I = 1}} = \left( {\begin{array}{*{20}{c}}
	  q \\ 
	  {\tilde q} 
	\end{array}} \right),\;\;\;\;{q_{I = 2}} = \left( {\begin{array}{*{20}{c}}
	  { - {{\tilde q}^\dag }} \\ 
	  {{q^\dag }} 
	\end{array}} \right)\;.
\end{equation}
The $A=1$ and $A=2$ components reside in complex conjugate representations $\mathfrak R, \bar {\mathfrak R}$ of the gauge and/or flavor group $G$. A hermitian generator $T \in \mathfrak g$ in representation $\mathfrak R$ acts on any field $K_{A}$ as
\begin{equation}
	T \cdot {K_{A}} = \left( {\begin{array}{*{20}{c}}
	  {T K_{1}} \\ 
	  {-T^* K_{2}} 
	\end{array}} \right)\;,
\end{equation}
and thus an adjoint field $\Xi = {\Xi^a}{T^a}$ acts as
\begin{equation}
	\Xi  \cdot K_{A} = {\Xi ^a}{T^a} \cdot K_A = \left( {\begin{array}{*{20}{c}}
	  {{\Xi ^a}{T^a}}&0 \\ 
	  0&{ - {\Xi ^a}({T^a})^* } 
	\end{array}} \right)\left( {\begin{array}{*{20}{c}}
	  {{K_{A = 1}}} \\ 
	  {{K_{A = 2}}} 
	\end{array}} \right)\;.
\end{equation}

The supersymmetry transformation rules are \cite{Hama:2012bg}
\begin{equation}
	\left\{ \begin{gathered}
	 \mathcal  Qq_{IA} =  - i\left( {{\xi _I}{\psi_A}} \right) + i({{\tilde \xi }_I}{{\tilde \psi }_A}) \hfill \\
	 \mathcal  Q{\psi _A} =  - 2{\sigma ^\mu }{{\tilde \xi }^I}{D_\mu }q_{IA} - {\sigma ^\mu }{D_\mu }{{\tilde \xi }^I}q_{IA} + 4i \xi ^I\tilde \phi \cdot q_{IA} - 2{{\zeta }^{I^\prime}}F_{I^\prime A} \hfill \\
	\mathcal   Q{{\tilde \psi }_A} =  - 2{{\tilde \sigma }^\mu }{\xi ^I}{D_\mu }q_{IA} - {{\tilde \sigma }^\mu }{D_\mu }{\xi ^I}q_{IA} + 4i \tilde \xi^I \phi \cdot q_{IA} - 2{{\tilde \zeta }^{I^\prime}}F_{I^\prime A} \hfill \\
	\mathcal   QF_{I^\prime A} = i({{\zeta }_{I^\prime}}{\sigma ^\mu }{D_\mu }{{\tilde \psi }_A}) - 2 ({{\zeta }_{I^\prime}}\phi\cdot{\psi _A}) - 2({{\zeta }_{I^\prime}}{\lambda _J}){q^{J}_{A}} + 2i{T_{\mu \nu }}({{\zeta }_{I^\prime}}{\sigma ^{\mu \nu }}{\psi _A}) \hfill \\
	\qquad\qquad - i(\tilde \zeta_{I^\prime} {{\tilde \sigma }^\mu }{D_\mu }{\psi _A}) + 2 ({{\tilde \zeta_{I^\prime} }}\tilde\phi\cdot{{\tilde \psi }_A}) + 2({{\tilde \zeta_{I^\prime}}}{{\tilde \lambda }_J}){q^{J}_{A}} - 2i{{\tilde T}_{\mu \nu }}(\tilde \zeta_{I^\prime} {{\tilde \sigma }^{\mu \nu }}{{\tilde \psi }_A})\;. \hfill \\ 
	\end{gathered}  \right.
	\label{hyper-multiplet-transformation}
\end{equation}
where the extra symplectic-Majorana spinors $\zeta$ and $\tilde \zeta$ satisfy
\begin{equation}\label{zetaproperties}
	({\xi _I}{\zeta _{I^\prime}}) = ({{\tilde \xi }_I}{{\tilde \zeta }_{I^\prime}})\;,\qquad ({{\tilde \zeta }_{I^\prime}}{{\tilde \zeta }^{I^\prime}}) = s\;,\qquad ({\zeta ^{I^\prime}}{\zeta _{I^\prime}}) = \tilde s\;,\qquad {R^\mu } + ({\zeta ^{I^\prime}}{\sigma ^\mu }{{\tilde \zeta }_{I^\prime}}) = 0\;.
\end{equation}

The supersymmetric action for the hypermultiplet is $\mathcal Q$-exact \cite{Hama:2012bg} on the ellipsoid and thus does not play a role in the localization computations of this paper. One should remark though that it is only positive definite upon choosing the alternative reality properties for the auxiliary fields
\begin{equation}\label{realityF}
\overline{F_{IA}} = - \epsilon^{IJ}\Omega^{AB} F_{JB}\;.
\end{equation}

\paragraph{Supersymmetry algebra.}
The supersymmetry algebra takes the form
\begin{equation}\label{susyalgebra}
\mathcal Q^2 = -2i \mathcal L_R^{A+V+\check V} + \text{Gauge}(\Phi) + \text{Scale}(w) + \text{R}_{U(1)_r}(\Theta) + \text{R}_{SU(2)_\mathcal{R}}\left(\Theta_{IJ}^{SU(2)_\mathcal{R}}\right) + \check{\text{R}}_{SU(2)_{\mathcal{R}^\prime}}\left(\check{\Theta}^{SU(2)_{\mathcal{R}^\prime}}_{I^\prime J^\prime}\right)\;,
\end{equation}
where $\mathcal L_R^{A+V+\check V}$ denotes a gauge, $SU(2)_\mathcal{R}$ and $SU(2)_{\mathcal{R}^\prime}$-covariant Lie derivative along the vector field $R$, and $SU(2)_{\mathcal R^\prime}$ rotates the hypermultiplet auxiliary fields. The parameters are given by
\begin{align}
R^\mu  &\equiv ({\xi^I}{\sigma^\mu }{{\tilde \xi }_I})\;, \\
\Phi &\equiv 2i\phi \tilde s + 2i\tilde \phi s\;, \\
w &\equiv -2\left((\xi^I \tilde \xi^\prime_I)+(\tilde\xi_I \tilde \xi^{\prime I})\right)\;,\\
\Theta &\equiv -\left( (\xi^I \tilde \xi^\prime_I)-(\tilde\xi_I \tilde \xi^{\prime I}) \right)\;,\\
\Theta^{SU(2)_\mathcal{R}}_{IJ} &\equiv -4\left( (\xi_{(I}\xi^\prime_{J)}) - (\tilde \xi_{(I} \tilde \xi^\prime_{J)}) \right)\;,\\
\check\Theta^{SU(2)_{\mathcal{R}^\prime}}_{I^\prime J^\prime} &\equiv 2 i (\zeta_{(I^\prime}\sigma^\mu D_\mu \tilde\zeta_{J^\prime)})-2 i (D_\mu\zeta_{(I^\prime}\sigma^\mu \tilde\zeta_{J^\prime)}) + 4 i (\zeta_{(I^\prime}\sigma^{kl}T_{kl}\zeta_{J^\prime)})- 4 i (\tilde\zeta_{(I^\prime}\tilde\sigma^{kl}\tilde T_{kl}\tilde\zeta_{J^\prime)})\;.
\end{align}

As in \cite{Hama:2012bg}, we restrict ourselves to Killing spinors $\xi_I, \tilde \xi_I$ such that no scale or $U(1)_r$ transformations appear in $\mathcal Q^2.$ The conditions 
\begin{equation}\label{orthocondition}
(\xi^I \tilde \xi^\prime_I)=(\tilde\xi_I \tilde \xi^{\prime I})=0
\end{equation} 
can be solved for $\xi'_I$ and $\tilde \xi'_I$ as
\begin{equation}\label{Stensor}
	{{\xi'_I}} =  - i{S_{\mu \nu }}{\sigma ^{\mu \nu }}{\xi _I}\;,\qquad {{\tilde \xi '}_I} =  - i{{\tilde S}_{\mu \nu }}{{\tilde \sigma }^{\mu \nu }}{{\tilde \xi }_I}\;.
\end{equation}

\section{Killing Spinors And Complex Structures}\label{appC}
In this subsection we introduce almost complex structures whose existence is guaranteed by having a solution to the generalized Killing spinor equations \eqref{Killing-spinor-equations} and the auxiliary equations \eqref{auxiliary-equations}. They will turn out to be useful when analyzing the singular solutions to the BPS equations.

The interplay between supersymmetry and geometry is quite rich, as for example observed for four-dimensional theories with four or fewer supercharges in \cite{Festuccia:2011ws,Dumitrescu:2012ha,Dumitrescu:2012at,Closset:2013vra,Closset:2014uda}. It is clear that we are only scratching the surface here, and a more in depth analysis would be very interesting.

\subsection{Locally Almost Complex Structures}
\label{appendix:almost-complex-structures}

Let $\xi_I$ and $\tilde \xi_I$ to be the solutions to the generalized Killing spinor equations (\ref{Killing-spinor-equations}) and the auxiliary equations (\ref{auxiliary-equations}). Then given any symplectic-Majorana spinor $\chi_I$ and $\tilde \chi_I$ such that $({\xi _I}{\chi ^I}) = 0$, $({\xi _I}{\chi ^I}) = 0$, one can define two almost complex structures away from the zeros of $\xi$ and $\tilde \xi$
\begin{equation}
{J^\mu }_\nu  \equiv \frac{1}{{\sqrt {s{s_\chi }} }}\left( {{\xi _I}{\sigma ^\mu }_\nu {\chi ^I}} \right)\;, \qquad {{\tilde J}^\mu }{_\nu}  \equiv \frac{1}{{\sqrt {\tilde s{s_{\tilde \chi }}} }}({{\tilde \xi }_I}{\tilde \sigma ^\mu }{_\nu} {{\tilde \chi }^I})\;,
	\label{definition:almost-complex-structures}
\end{equation}
where ${s_\chi } \equiv \left( {{\chi ^I}{\chi _I}} \right)$, ${s_{\tilde \chi }} \equiv ({{\tilde \chi }_I}{{\tilde \chi }^I})$ are both positive semi-definite. Using Fierz identities, it is easy to check that
\begin{equation}
	{J^\mu }_\lambda {J^\lambda }_\nu  =  - \delta _\nu ^\mu\;, \qquad {{\tilde J}^\mu }{_\lambda }{{\tilde J}^\lambda }{_\nu } =  - \delta _\nu ^\mu\;, \qquad {J^\mu }_\lambda {{\tilde J}^\lambda }{_\nu}  = {{\tilde J}^\mu }{_\lambda} {J^\lambda }_\nu \;.
\end{equation}

Where $\xi$ is non-zero, one can write ${\chi _I} = {m_I}^J{\xi _J}$ or ${\chi _I} = {m_{\mu \nu }}{\sigma ^{\mu \nu }}{\xi _I}$ and similarly for $\tilde \chi$, where $m_{IJ}$ is a triplet of functions satisfying $\overline {{m_{IJ}}}  = {\varepsilon ^{II'}}{\varepsilon ^{JJ'}}{m_{I'J'}}$, and $m_{\mu \nu}$ is a real anti-self-dual 2-form. The two representations are interchangeable, for instance, $m_{IJ}$ and $m_{\mu \nu}$ are related by
\begin{equation}
	{m_{\mu \nu }}\Theta _{IJ}^{\mu \nu } = \frac{1}{2}s {m_{IJ}}.
\end{equation}
In the following, when we need to, we pick the representation of $\chi_I$ using $m_{IJ}$, and similarly for $\tilde \chi_I$.

On open sets where the locally almost complex structures are defined, one can introduce the decomposition of tangent vectors with respect to $J$ and $\tilde J$ respectively:
\begin{equation}
	\left\{ \begin{gathered}
	  J{X^{1,0}} = i{X^{1,0}} \hfill \\
	  J{X^{0,1}} =  - i{X^{0,1}} \hfill \\ 
	\end{gathered}  \right.,\;\;\;\;\left\{ \begin{gathered}
	  \tilde J{{\tilde X}^{1,0}} = i{{\tilde X}^{1,0}} \hfill \\
	  \tilde J{X^{0,1}} =  - i{{\tilde X}^{0,1}} \hfill \\ 
	\end{gathered}  \right.
\end{equation}

A $(p,q)$-type vector can be characterized using spinorial equations. First of all, taking $JX = iX$ as an example,
\begin{equation}
		JX = iX \Leftrightarrow \left\{ \begin{gathered}
	  \left[ {\frac{1}{{\sqrt {s{s_\chi }} }}\left( {{\xi ^I}{\sigma ^\mu }{{\tilde \sigma }_\nu }{\chi _I}} \right) + i\frac{1}{s}\left( {{\xi ^I}{\sigma ^\mu }{{\tilde \sigma }_\nu }{\xi _I}} \right)} \right]{X^\nu } = 0 \hfill \\
	  \left[ {\frac{1}{{\sqrt {s{s_\chi }} }}\left( {{\xi ^I}{\sigma ^\mu }{{\tilde \sigma }_\nu }{\chi _I}} \right) + i\frac{1}{{{s_\chi }}}\left( {{\chi ^I}{\sigma ^\mu }{{\tilde \sigma }_\nu }{\chi _I}} \right)} \right]{X^\nu } = 0 \hfill \\ 
	\end{gathered}  \right..
\end{equation}
Multiplying $\overline X^\mu$ to the two equations on the right, and subsequently taking their sum, it is easy to verify that one obtains a semi-positive product
\begin{equation}
\sum\limits_{I,\dot \alpha } {\Delta _I^{\dot \alpha }\overline {\Delta _I^{\dot \alpha }} }  = \overline {{X^\mu }} \left[ {s_\chi ^{ - 1}\left( {{\chi ^I}{\sigma _{\mu \nu }}{\chi _I}} \right) + 2i{s^{ - 1/2}}s_\chi ^{ - 1/2}\left( {{\chi ^I}{\sigma _{\mu \nu }}{\xi _I}} \right) + {s^{ - 1}}\left( {{\xi ^I}{\sigma _{\mu \nu }}{\xi _I}} \right)} \right]{X^\nu }   \geqslant 0 \;,
\end{equation}
where $\Delta _I^{\dot \alpha } = {X^\mu }{({{\tilde \sigma }_\mu })^{\dot \alpha \gamma }}\left( {s_\chi ^{ - 1/2}{\chi _{I\gamma }} + i{s^{ - 1/2}}{\xi _{I\gamma }}} \right)$. Therefore,
\begin{equation}
	JX = iX \Leftrightarrow {\Delta _I} \equiv {X^\mu }{{\tilde \sigma }_\mu }\left( {s_\chi ^{ - 1/2}{\chi _I} + i{s^{ - 1/2}}{\xi _I}} \right) = 0\;.
\end{equation}
Similarly, one can derive the spinorial condition for $X$ to be $(p,q)$-vector of $J$ or $\tilde J$:
\begin{equation}
	\left\{ \begin{gathered}
	  JX = iX \Leftrightarrow {X^\mu }{{\tilde \sigma }_\mu }\left( {s_\chi ^{ - 1/2}{\chi _I} + i{s^{ - 1/2}}{\xi _I}} \right) = 0 \hfill \\
	  JX =  - iX \Leftrightarrow {X^\mu }{{\tilde \sigma }_\mu }\left( {s_\chi ^{ - 1/2}{\chi _I} - i{s^{ - 1/2}}{\xi _I}} \right) = 0 \hfill \\ 
	\end{gathered}  \right.,
\end{equation}
\begin{equation}
	\left\{ \begin{gathered}
	  \tilde JX = iX \Leftrightarrow {X^\nu }{\sigma _\nu }\left( {\tilde s_{\tilde \chi }^{ - 1/2}{{\tilde \chi }_I} - i{s^{ - 1/2}}{{\tilde \xi }_I}} \right) = 0 \hfill \\
	  \tilde JX =  - iX \Leftrightarrow {X^\nu }{\sigma _\nu }\left( {\tilde s_{\tilde \chi }^{ - 1/2}{{\tilde \chi }_I} + i{s^{ - 1/2}}{{\tilde \xi }_I}} \right) = 0 \hfill \\ 
	\end{gathered}  \right..
\end{equation}

\subsection{Integrability}

It is possible that the almost complex structures induced by Killing spinors are integrable. In the following, we consider $\chi_I = {m_I}^J \xi_J$ and study the conditions for $J$ to be integrable.

Before moving on to the detail, let us make a remark. Notice that $J_{\mu\nu}$ is anti-self-dual and $\tilde J_{\mu\nu}$ is self-dual. That implies that, with the implicitly chosen orientation, one has the decomposition of self-dual and anti-self-dual 2-forms
\begin{equation}
	\left\{ \begin{gathered}
	  {\Omega ^ + } =  \Omega _{\tilde J}^{2,0} \oplus \Omega _{\tilde J}^{0,2} \oplus \tilde J \cdot {\Omega ^0} = \Omega_J^{\prime 1,1} \hfill \\
	  {\Omega ^ - } = \Omega _J^{2,0} \oplus \Omega _J^{2,0} \oplus J \cdot {\Omega ^0} = \Omega _{\tilde J}^{\prime 1,1} \hfill \\ 
	\end{gathered}  \right.\;,
\end{equation}
where the prime indicates removing components along $J$ and $\tilde J$ in the first and second line respectively. Therefore, for instance, if $X$, $Y$ are $(1,0)_J$ vectors, then any self-dual 2-form $\tilde \omega$ satisfies
\begin{equation}
	{{\tilde \omega}_{\mu \nu }}\left( {X,Y} \right) = 0
\end{equation}
because $\tilde \omega$ has no components in $\Omega^{2,0}_J$.

Let $X$, $Y$ be $(1,0)_J$ vectors with respect to $J$. We wish to analyze which conditions guarantee that their Lie bracket is still of type $(1,0)_J,$ \ie{}, we want to study when
\begin{equation}
	\left( {{X^\mu }{\nabla _\mu }{Y^\nu } - {Y^\mu }{\nabla _\mu }{X^\nu }} \right){\tilde \sigma _\nu }\left( {s_\chi ^{ - 1/2}{\chi _I} + i{s^{ - 1/2}}{\xi _I}} \right) = 0\;.
\end{equation}
Note that ${s_\chi } = \frac{1}{2}s{m^{IJ}}{m_{IJ}} \equiv s \mathcal{M}$, hence the condition can be rewritten as
\begin{align}
0 &= \left( {{X^\mu }{Y^\nu } - {X^\nu }{Y^\mu }} \right){{\tilde \sigma }_\nu }{D_\mu }\left( {{{\mathcal M}^{ - 1/2}}{m_I}^J{\xi _J} + i{\xi _I}} \right) \\
&= \left( {{X^\mu }{Y^\nu } - {X^\nu }{Y^\mu }} \right){{\tilde \sigma }_\nu }\left( {{D_\mu }\left( {{{\mathcal M}^{ - 1/2}}{m_I}^J} \right){\xi _J} + {{\mathcal M}^{ - 1/2}}{m_I}^J{D_\mu }{\xi _J} + i{D_\mu }{\xi _I}} \right) \\
& = \left( {{X^\mu }{Y^\nu } - {X^\nu }{Y^\mu }} \right){{\tilde \sigma }_\nu } {{D_\mu }\left( {{{\mathcal M}^{ - 1/2}}{m_I}^J} \right){\xi _J}} \nn\\
& \qquad  - \left( {{X^\mu }{Y^\nu } - {X^\nu }{Y^\mu }} \right)\left({{\mathcal M}^{ - 1/2}}{m_I}^J + i \delta_I^J\right)\left( {{T^{\lambda \rho }}{{\tilde \sigma }_\nu }{\sigma _{\lambda \rho }}{\sigma _\mu }{{\tilde \xi }_J} + i{{\tilde \sigma }_{\nu \mu }}{{\tilde \xi '}_J}} \right) \;.
\end{align}
Using the fact that a self-dual 2-form has no $(2,0)_J$ components and ${T^{\lambda \rho }}\left( {{{\tilde \sigma }_\mu }{\sigma _{\lambda \rho }}{\sigma _\nu } - \mu  \leftrightarrow \nu } \right) = 8{T_{\mu \nu }}$, the condition reduces to
\begin{equation}
	0 = \left( {{X^\mu }{Y^\nu } - {X^\nu }{Y^\mu }} \right)\left[ {{{\tilde \sigma }_\nu }{D_\mu }({{\mathcal M}^{ - 1/2}}{m_I}^J){\xi _J} + {T_{\mu \nu }}({{\mathcal M}^{ - 1/2}}{m_I}^J{{\tilde \xi }_J} + i{{\tilde \xi }_I})} \right]\;.
\end{equation}
This expression vanishes and therefore $J$ is integrable if
\begin{equation}\label{integrabilitycondition}
{D_\mu }\left( {\frac{1}{{\sqrt {{m_{KL}}{m^{KL}}} }}{m_I}^J} \right) = 0\;,\qquad {T_J^{2,0}} = {T_J^{0,2}} = 0\;.
\end{equation}


\section{Geometry of Ellipsoid }\label{appendix:ellipsoid}
\subsection{Some useful properties}
Let us first list a few useful general properties of the bilinears \eqref{bilinearsofKS} one can construct given a Killing spinor solution $\xi_I$ and $\tilde \xi_I$. As a direct consequence of \eqref{Killing-spinor-equations}, one finds
\begin{itemize}
	\item ${\partial _\mu }s = 8{R^\nu }({T_{\mu \nu }} - {{\tilde S}_{\mu \nu }})$, ${\partial _\mu }\tilde s = 8{R^\nu }({{\tilde T}_{\mu \nu }} - {S_{\mu \nu }})$.
	\item $d{\kappa _{\mu \nu }} = 8\tilde s({T_{\mu \nu }} - {{\tilde S}_{\mu \nu }}) + 8s({{\tilde T}_{\mu \nu }} - {S_{\mu \nu }})$, which implies ${R^\mu }d{\kappa _{\mu \nu }} = {\partial _\mu }(s\tilde s)\label{d-kappa}$.
\end{itemize}
Following \cite{Hama:2012bg}, we define
\begin{equation}
	{w_{IJ}} \equiv 4{s^{ - 1}}\Theta _{IJ}^{\mu \nu }({T_{\mu \nu }} - {S_{\mu \nu }}),\;\;\;\;{\tilde w_{IJ}} \equiv  - 4{\tilde s^{ - 1}}\tilde \Theta _{IJ}^{\mu \nu }({\tilde T_{\mu \nu }} - {\tilde S_{\mu \nu }})\;.
	\label{redefinition:w_IJ}
\end{equation}
By a Fierz identity, $w_{IJ}$ and $\tilde w_{IJ}$ satisfy
\begin{equation}\label{propertyw_IJ}
	{w_{IJ}}{\xi ^J} =  - ({T_{\mu \nu }} - {S_{\mu \nu }}){\sigma ^{\mu \nu }}{\xi _I},\;\;\;\;,\;\;\;\;{{\tilde w}_{IJ}}{{\tilde \xi }^J} =  - ({{\tilde T}_{\mu \nu }} - {{\tilde S}_{\mu \nu }}){{\tilde \sigma }^{\mu \nu }}{{\tilde \xi }_I}\;.
\end{equation}

\subsection{Ellipsoid}
The ellipsoid can be defined by its embedding equation in $\mathbb R^5$
\begin{equation}\label{ellipsoidembedding}
\frac{x_0^2}{r^2} + \frac{x_1^2 + x_2^2}{\ell^2} + \frac{x_3^2+x_4^2}{\tilde \ell^2} = 1\;.
\end{equation}
Introducing polar coordinates
\begin{equation}\label{embeddingcoordinates}
\begin{aligned}
x_0 &= r \cos\rho\;, \qquad &x_1 &= \ell \sin \rho \ \cos\theta\ \cos\varphi\;, \qquad &x_3 &= \tilde\ell \sin \rho \ \sin\theta\ \cos\chi\;, \\
& &x_2 &= \ell \sin \rho \ \cos\theta\ \sin\varphi\;, \qquad &x_4 &= \tilde\ell \sin \rho \ \sin\theta\ \sin\chi\;,
\end{aligned}
\end{equation}
its metric can be written in terms of the vielbeins
\begin{equation}
e^1 = \ell \sin\rho \cos\theta\  d\varphi\;, \quad e^2 = \tilde\ell \sin\rho \sin\theta\  d\chi\;, \quad e^3 = f \sin\rho\ d\theta + h\  d\rho\;, \quad e^4 = g\ d\rho\;,
\end{equation}
where $f=\sqrt{\ell^2 \sin^2\theta + \tilde \ell^2 \cos^2\theta},$ $g=\sqrt{r^2 \sin^2 \rho + \ell^2\tilde \ell^2 f^{-2}\cos^2\rho},$ and $h = \frac{\tilde \ell^2 - \ell^2}{f} \cos\rho\sin\theta\cos\theta.$

In \cite{Hama:2012bg}, a solution to the generalized Killing spinor equations \eqref{Killing-spinor-equations} and the auxiliary equations \eqref{auxiliary-equations}, also satisfying the orthogonality condition \eqref{orthocondition}, was found. It reads
\begin{equation}
	\left\{ \begin{gathered}
	  {\xi _{I = 1}} = \frac{1}{2}\sin \frac{\rho }{2}\left( {\begin{array}{*{20}{c}}
	  {{e^{i\left( {\chi  + \varphi  - \theta } \right)/2}}} \\ 
	  { - {e^{i\left( {\chi  + \varphi  + \theta } \right)/2}}} 
	\end{array}} \right) \hfill \\
	  {\xi _{I = 2}} = \frac{1}{2}\sin \frac{\rho }{2}\left( {\begin{array}{*{20}{c}}
	  {{e^{i\left( { - \chi  - \varphi  - \theta } \right)/2}}} \\ 
	  {{e^{i\left( { - \chi  - \varphi  + \theta } \right)/2}}} 
	\end{array}} \right) \hfill \\ 
	\end{gathered}  \right.,\qquad \left\{ \begin{gathered}
	  {{\tilde \xi }_{I = 1}} = \frac{i}{2}\cos \frac{\rho }{2}\left( {\begin{array}{*{20}{c}}
	  {{e^{i\left( {\chi  + \varphi  - \theta } \right)/2}}} \\ 
	  { - {e^{i\left( {\chi  + \varphi  + \theta } \right)/2}}} 
	\end{array}} \right) \hfill \\
	  {{\tilde \xi }_{I = 2}} =  - \frac{i}{2}\cos \frac{\rho }{2}\left( {\begin{array}{*{20}{c}}
	  {{e^{i\left( { - \chi  - \varphi  - \theta } \right)/2}}} \\ 
	  {{e^{i\left( { - \chi  - \varphi  + \theta } \right)/2}}} 
	\end{array}} \right) \hfill \\ 
	\end{gathered}  \right.\;.
\end{equation}
The corresponding explicit expressions for the auxiliary fields $T_{\mu\nu},\tilde T_{\mu\nu},S_{\mu\nu},\tilde S_{\mu\nu}, V_{\mu}$ and $M$ can be found in \cite{Hama:2012bg}.

Introducing ${\tau _\theta } \equiv \cos \theta {\tau _1} + \sin \theta {\tau _2}$ and $\tau_{2,\theta} \equiv i{\tau _\theta }{\tau _3},$ one can note that $\xi_I$ and $\tilde \xi_I$ are eigenvectors of $\tau_\theta$: ${\tau _\theta }{\xi _1} =  - {\xi _1}$, ${\tau _\theta }{\xi _2} = {\xi _2}$, ${\tau _\theta }{\tilde \xi _1} =  - {\tilde \xi _1}$, ${\tau _\theta }{\tilde \xi _2} = {\tilde \xi _2}	$.
Furthermore, one finds for the simplest bilinears defined in \eqref{bilinearsofKS}
\begin{equation}
s \equiv (\xi^I \xi_I) = {\sin ^2}\frac{\rho }{2}\;, \qquad \tilde s \equiv (\tilde \xi_I \tilde \xi^I) = {\cos ^2}\frac{\rho }{2}\;, \qquad {R^\mu } \equiv ({\xi ^I}{\sigma ^\mu }{\tilde \xi _I}) = - \frac{{\sin \rho }}{2}\left( {\cos \theta e_1^\mu  + \sin \theta e_2^\mu } \right)\;.
	\label{equation:bilinears-on-ellipsoid}
\end{equation}
In particular one finds that $s+\tilde s = 1.$ It is also important to note that $w_{IJ}$ and $\tilde w_{IJ}$ defined in \eqref{redefinition:w_IJ} are equal on the ellipsoid and using \eqref{propertyw_IJ} can thus be used to define a Fayet-Iliopoulos action as in \eqref{FIaction}.

As discussed in appendix \ref{appendix:almost-complex-structures}, one can define various almost complex structures away from the north or south poles. In particular, we consider
\begin{equation}
{J^\mu }_\nu  \equiv 2i{s^{ - 1}}{\left( {{\Theta _{12}}} \right)^\mu }_\nu\;, \qquad {\tilde J^\mu }{_\nu}  \equiv 2i{{\tilde s}^{ - 1}}{({{\tilde \Theta }_{12}})^\mu }_\nu\;.
	\label{definition:complex-structures}
\end{equation}
where $J$ is defined away from the north pole ($\rho = 0$), and $\tilde J$ is defined away from the south pole. Here we chose $\chi_I\propto (\tau^3)_I^{\phantom{I}J} \xi_J,$ and $\tilde \chi_I \propto (\tau^3)_I^{\phantom{I}J} \tilde\xi_J.$ In vielbein indices, the two almost complex structures read
\begin{equation}
J = \left( {\begin{array}{*{20}{c}}
	  0&0&{ - \sin \theta }&{ - \cos \theta } \\ 
	  0&0&{\cos \theta }&{ - \sin \theta } \\ 
	  {\sin \theta }&{ - \cos \theta }&0&0 \\ 
	  {\cos \theta }&{\sin \theta }&0&0 
	\end{array}} \right)\;, \qquad \tilde J = \left( {\begin{array}{*{20}{c}}
	  0&0&{ - \sin \theta }&{\cos \theta } \\ 
	  0&0&{\cos \theta }&{\sin \theta } \\ 
	  {\sin \theta }&{ - \cos \theta }&0&0 \\ 
	  { - \cos \theta }&{ - \sin \theta }&0&0 
	\end{array}} \right)\;.
\end{equation}
One can verify that the conditions \eqref{integrabilitycondition} are satisfied and thus that they are integrable.

Note also that the forms $\Theta_{IJ}$ and $\tilde \Theta_{IJ}$ for equal indices $I=J$ are elements of the $(2,0)$ or $(0,2)$-forms with respect to $J$ and $\tilde J$:
\begin{equation}
	{\Theta _{11}} \in \Omega _J^{0,2},\;\;\;\;{\Theta _{22}} \in \Omega _J^{2,0},\;\;\;\;{\tilde \Theta _{11}} \in \Omega _{\tilde J}^{0,2},\;\;\;\;{\tilde \Theta _{22}} \in \Omega _{\tilde J}^{2,0}
\end{equation}
Near $\rho = 0$, the complex structure $\tilde J$ reduces to the opposite of the usual complex structure on $\mathbb{R}^4 = \mathbb{C} ^ 2$ parameterized by $(x_1, x_2, x_3, x_4)$, and $T_{\tilde J}^{1,0} = \operatorname{span}\left\{ {{\partial _{{x_1}}} + i{\partial _{{x_2}}},{\partial _{{x_3}}} + i{\partial _{{x_4}}}} \right\}$.

\subsection{Round $S^4$}
Setting $\ell = \tilde \ell = r,$ the ellipsoid in \eqref{ellipsoidembedding} becomes the round sphere of radius $\ell.$ On this round geometry, various simplifications occur and some special properties help simplify the discussion. First of all, some of the auxiliary fields appearing in the generalized Killing spinor equations and the auxiliary equations vanish:
\begin{equation}
{T_{\mu \nu }} = {\tilde T_{\mu \nu }} = {\left( {{V_\mu }} \right)_I}^J = 0\;,
\end{equation}
while the auxiliary field $M$ is simply given by
\begin{equation}
M = -\frac{4}{\ell^2}\;.
\end{equation}
For $\xi'_I = -i S_{\mu\nu}\sigma^{\mu\nu}\xi_I$ and $\tilde \xi'_I = -i \tilde S_{\mu\nu}\tilde\sigma^{\mu\nu}\tilde\xi_I$ one finds
\begin{equation}
	{\xi'_1} = \frac{1}{{2\ell }}{\xi _1}\;, \qquad {\xi'_2} =  - \frac{1}{{2\ell }}{\xi _2}\;, \qquad {\tilde \xi'_1} = \frac{1}{{2\ell }}{\tilde \xi _1}\;, \qquad {\tilde \xi '_2} =  - \frac{1}{{2\ell }}{\tilde \xi _2}\;.
\end{equation}

Recalling that $d\kappa$ can be expanded as in \eqref{d-kappa}, one finds on the round sphere, $d{\kappa _{\mu \nu }} =  - 8\tilde s{\tilde S_{\mu \nu }} - 8s{S_{\mu \nu }}$. Hence one has
\begin{equation}
	\begin{gathered}
	  d\kappa _{\mu \nu }^ -  =  - 8s{S_{\mu \nu }} = s{\ell ^{ - 1}}{J_{\mu \nu }} = 2i{\ell ^{ - 1}}{({\Theta _{12}})_{\mu \nu }}, \hfill \\
	  d\kappa _{\mu \nu }^ +  =  - 8\tilde s{{\tilde S}_{\mu \nu }} = - \tilde s{\ell ^{ - 1}}{{\tilde J}_{\mu \nu }} =  - 2 i {\ell ^{ - 1}}{({{\tilde \Theta }_{12}})_{\mu \nu }}. \hfill \\ 
	\end{gathered} 
\end{equation}
and ${R^\mu }({S_{\mu \nu }} + {{\tilde S}_{\mu \nu }}) = 0$.

Finally, the triplet of functions $w_{IJ}=\tilde w_{IJ}$ defined in \eqref{redefinition:w_IJ} read:
\begin{equation}
	{w_{12}} = {w_{21}} = {{\tilde w}_{12}} = {{\tilde w}_{21}} = \frac{1}{{i\ell }}\;, \qquad {w_{11}} = {w_{22}} = {{\tilde w}_{11}} = {{\tilde w}_{22}} = 0\;.
\end{equation}


{
\bibliographystyle{utphys}
\bibliography{HBLS4_draft}
}

\end{document}